\documentclass{JHEP3}
\usepackage{graphicx}
\usepackage{epsfig}

\usepackage{bm}
\usepackage{amssymb}
\usepackage{amsmath}

\def\fun#1#2{\lower3.6pt\vbox{\baselineskip0pt\lineskip.9pt
  \ialign{$\mathsurround=0pt#1\hfil##\hfil$\crcr#2\crcr\sim\crcr}}}

\def\fun#1#2{\lower3.6pt\vbox{\baselineskip0pt\lineskip.9pt
\ialign{$\mathsurround=0pt#1\hfil##\hfil$\crcr#2\crcr\sim\crcr}}}

\def\mpl{m_{\rm Pl}}
\newcommand{\MUNCH}[1]{\relax}

\title{Recovering the Inflationary Potential and Primordial Power Spectrum With a Slow Roll Prior: Methodology and Application 
to WMAP 3 Year Data}

\author{Hiranya V. Peiris\thanks{Hubble Fellow}  \\
Kavli Institute for Cosmological Physics and Enrico Fermi Institute, \\University of Chicago, Chicago IL 60637, USA. \\
\email{hiranya@cfcp.uchicago.edu}}

 \author{Richard Easther \\
Department of Physics, Yale University, New Haven CT 06520, USA \\
\email{richard.easther@yale.edu}}

\abstract{ We introduce a new method for applying an inflationary prior to a cosmological dataset that includes relations between observables at arbitrary order in the slow roll expansion.  The process is based on the inflationary flow equations, and the slow roll parameters appear explicitly in the cosmological parameter set.  We contrast our method to other ways of imposing an inflationary prior on a cosmological dataset, and argue that this method is ideal for use with heterogeneous datasets. In particular, it would be well suited to exploiting any direct detection of fundamental tensor modes by a BBO-style mission.  To demonstrate the practical use of this method we apply it to the  WMAPI+All dataset, and the newly released WMAPII dataset on its own and together with the SDSS data.  We find that all basic classes of single field inflationary models are still allowed at the $1-2\sigma$ level, but the overall parameter  space is sharply constrained.   In particular, we find evidence that the combination  of WMAPII+SDSS is sensitive to effects arising from terms that are quadratic in the two leading-order slow roll parameters. 
}

\begin{document}

\section{Introduction}

Observational cosmology is a booming science.   We typically see at least one significant data release each  year, from both Cosmic Microwave Background [CMB] experiments and probes of Large Scale Structure [LSS]. Several  generations of experiments are now being designed and constructed, suggesting that current rates of progress will continue well into the future.  A major and very welcome challenge for the cosmological community is how to best exploit this torrent of data. Each major data release is typically accompanied by newly estimated values of the standard cosmological parameter set -- based either on the new dataset alone, or an updated  ``world average'' drawn from the combination of  several, complementary experiments.  

There are two ways to construct a cosmological parameter set -- the first and more general is to use a purely empirical set of observables which contains as few theoretical assumptions as possible. Alternatively, one may impose a ``prior'' corresponding to the predictions of a specific theoretical model.  Because this reduces the overall parameter space, this class of variables can be more tightly constrained than a generic set.  The most common theoretical prior is derived from inflation.   This typically puts two cuts on the parameter set. Firstly, without careful tuning, inflation predicts that present day universe is flat, so we impose $\Omega_{\mbox{\tiny Total}} = 1$. Secondly, generic models of single field, slow roll inflation obey a {\em consistency condition\/} that   relates   the amplitudes of the primordial scalar and tensor spectra with the slope of the tensor spectrum (see \cite{lidsey/etal:1995} for a discussion.)   In addition to these two constraints, the conventional description of the primordial spectra in terms of their amplitude $A$ and spectral index $n$  is unnecessary, if one assumes that the primordial perturbations  are inflationary.\footnote{As opposed to both ``defect'' models such as cosmic  strings, and also non-standard inflationary schemes, such as the curvaton scenario \cite{Lyth:2001nq}. In the latter case, inflation solves the ``global'' cosmological problems such as homogeneity, isotropy and flatness, but the perturbation spectrum is not immediately determined by the inflationary potential.}  

The spectral indices are derived from a truncated Taylor expansion of the overall power-spectrum,   $P(k)$, where $k$ is the comoving wavenumber.   One may extend this expansion to higher order by including the running, $\alpha = d n_s /d\ln{k}$, as an additional parameter.  This is obviously an empirical characterization of the spectrum. Inflationary theory  expresses  these parameters as  functions of the potential $V(\phi)$ and its  derivatives.    This introduces a  further truncated expansion -- this time in terms of the slow roll parameters, which are dimensionless combinations of derivatives of the potential.  This expansion is independent of the Taylor expansion underlying the spectral indices, and ignores the relationships between the parameters of the spectra found in the slow roll limit. Consequently, once we have adopted an inflationary prior there is nothing to be gained by expressing the spectrum in terms of $n_s$ and $d n_s /d\ln{k}$, and doing so necessarily invokes a further and ultimately unnecessary truncated expansion and risks discarding useful information. In this paper we will investigate the use of a set of cosmological parameters which incorporates the slow roll variables directly, making no direct use of the  spectral index or its running.   

If all we gained from this substitution was a reshuffling of the variables that describe the perturbation spectrum, the issue would  be of purely academic  interest. However, the slow roll parameters have an immediate physical meaning in inflationary models, and including them  in our parameter set raises a number of new and exciting possibilities. In conventional fits to the spectrum, one specifies the amplitude and index at some fiducial wavenumber, $k_0$ ($0.002  \mbox{Mpc}^{-1}$ in WMAP data), and the slow roll parameters are likewise measured at some fixed wavenumber.  The simplest way to implement this in a CMB code such as CAMB or CMBFast is to express $n_s$ and $d n_s/d\ln{k}$  as a function of the slow roll parameters at $k_0$, and then compute the CMB and LSS power spectra in the usual way.   However,  the slow roll parameters are not constant throughout the inflationary epoch but are fully specified by the inflationary flow equations \cite{hoffman/turner:2001,kinney:2002}, a set of coupled first order differential equations.  The values of the slow roll parameters at $k_0$ provide the initial data for this system, and  solving the flow equations yields values of the slow roll parameters at any other value of $k$.  Formally, the slow roll hierarchy is infinite, but any dataset will only contain enough information to put meaningful constraints on a finite subset of terms.  However, if we truncate the slow roll hierarchy at any given point, the structure of the equations ensures that the truncation holds everywhere, and does not depend explicitly on the fiducial wavenumber $k_0$.  

The  flow equations  contain  enough information for us to reconstruct the inflationary potential  up to an overall normalization -- which would only be fixed by an unambiguous measurement of the tensor contribution to the CMB.  Consequently, this  approach provides the most direct route to  a long-held ambition of early universe cosmology: recovering the inflationary potential directly from astrophysical data.   Previous attempts to do this have typically relied on computing the spectral parameters, inverting these to obtain the slow roll parameters that gave rise to them, and then solving these to obtain the underlying potential.    This process can be carried out analytically, as summarized in \cite{lidsey/etal:1995}.  To first order, the mappings between spectral indices and the potential slow roll parameters are relatively simple, but they become considerably more complicated at second order  \cite{Copeland:1993zn,Liddle:1994cr}. 
 By working directly with the Hubble slow roll hierarchy, our algorithm automatically incorporates any and all relations between the inflationary observables, to whatever order we choose to truncate the expansion.   In particular, we will see that the new WMAP dataset, in combination with SDSS data, has enough power to significantly constrain the allowed running of the scalar spectral index.  This arises from terms proportional to the square of the first two slow roll parameters, and is thus indicative of both the quality of present cosmological data, and the ability of our method to exploit this by including higher order relations between the slow roll parameters and inflationary observables.      
 
There are two important precedents for this work. The first is {\em Monte Carlo Reconstruction\/} \cite{easther/kinney:2003}, and the second is Leach {\it et al.\/}'s progress towards placing direct constraints of the slow roll parameters using both CMB and LSS data \cite{Leach:2002ar,Leach:2002dw,Leach:2003us}.   As initially proposed,  Monte Carlo reconstruction proceeds by solving the flow equations with arbitrary initial conditions and then identifying models  which match some predefined constraint(s) on the spectral parameters.  The problems with this approach is that existing implementations  a) consider a vast number of models which do not provide a good fit to data, b) there is no natural measure with which to weight the choice of initial values of the slow roll parameters and c) no attempt to establish a relative likelihood for the parameter values that do satisfy the chosen astrophysical constraints.   All these problems are solved when we obtain the slow roll parameters directly from a Monte Carlo Markov Chain [MCMC] fit to data.  One of the virtues of the MCMC approach is that one can draw parameter values from a flat distribution, and the adaptive nature of the chains ensures that you focus computational effort on the subspace where the likelihood is most sharply peaked.   Moreover, the relative merit of different parameter choices is quantified by the likelihood computed at each step in the chain with respect to the data.    

Conversely, Leach {\em et al.} \cite{Leach:2002ar,Leach:2002dw,Leach:2003us}, work with $n_s$ and $d n_s/d\ln{k}$,
as first suggested in \cite{Martin:2000ak}.   The cosmological observables are expressed in terms of the slow roll parameters, whose scale dependence is included via the slow roll approximation.
While this approach could, in principle, be extended to arbitrary order in slow roll, the algebraic complexity encountered at each successive order is significantly higher than the last \cite{Stewart:2001cd}.  On the other hand, because only a comparatively small piece of the inflationary potential is directly ``sampled'' by observations of the CMB and LSS, in most cases the second order results are sufficiently accurate for practical purposes. Indeed, in what follows, we compare our approach to that of  Leach {\em et al.} when  applied to the WMAPI+All dataset, and show that it leads to essentially identical constraints on the slow roll parameters.  On the other hand, the inevitable improvement of cosmological data will  increase both the range of wavelengths over which we can probe the spectrum, and the accuracy to which we can measure it. If one ever wishes to exceed the limitations of the second order expansion, then the flow equations provide an elegant route to reconstruction. In particular, if primordial gravitational waves are ever detected directly via a BBO-style experiment the range of scales over which we can access primordial perturbations would increase dramatically, and the flow equations would be immediately applicable to this scenario (see \cite{Ungarelli:2005qb,Boyle:2005se,Smith:2006xf,Chongchitnan:2006pe}).\footnote{BBO would detect inflationary perturbations at scales of a few million kilometers, rather than megaparsecs.}   In the absence of a direct detection of primordial gravitational waves at solar system scales, future data will still extend the range of $k$ values over which we can probe the primordial spectrum -- in particular, close quasar pairs facilitate short scale measurements of the Lyman-$\alpha$ forest power spectrum \cite{Lidz:2003fv}, and future measurements of the high redshift 21 centimeter radiation will probe the primordial power spectrum at  scales up to $k\sim10 \mbox{Mpc}^{-1}$  \cite{McQuinn:2005hk}.

A final advantage of this approach is that it allows us to impose constraints based on the number of e-folds that elapse between some fiducial mode with wavenumber $k_0$ crossing the horizon and the end of inflation.   In their work, Leach {\em et al.\/} note that because they extrapolate the power spectrum about a point there is a possibility that it will cross zero in the range of interest, leading to unphysical values for the power spectrum.  In our approach, we see that for some parameter values $\epsilon$ becomes larger than unity in the region of interest, signaling the end of inflation. In this case we can add a prior that inflation lasts longer than some minimal number of e-folds and exclude  the model on physical grounds.   

One of the virtues of the Hubble slow roll formalism is that, if it is used self consistently, the resulting inflationary dynamics actually provide an analytic solution to the evolution equations for a scalar field dominated FRW universe.   However,  the perturbation spectrum can  typically only be obtained exactly in very special circumstances \cite{Abbott:1984fp,Easther:1995pc,Starobinsky:2005ab}, so even with an exact result for the background one typically has to use the slow roll approximation in order to obtain the perturbation spectra.   Grivell and Liddle \cite{Grivell:1999wc} describe a reconstruction technique that again avoids the use of the spectral indices when computing inflationary perturbation spectra, and  goes one step further in that it entirely abandons the use of the slow roll approximation, directly integrating the fundamental mode equations in order to obtain the power spectrum.  Here we  employ slow roll expressions for the perturbation spectrum itself, accurate to second order in slow roll. This may appear paradoxical, but since we are accounting for the scale dependence in the slow roll parameters by using the flow equations, the higher order corrections to the potential are fed into $P_R$ and $P_h$ via the scale dependence of $\epsilon$ and $\eta$.  However, it would be entirely feasible to evaluate the spectrum exactly (up to the assumptions inherent in the use linear gravitational perturbation theory) in our approach.

Finally, we believe that using the HSR parameters directly in the chains avoids the theoretical uncertainties in the inflationary predictions associated with a given potential, discussed by Kinney and Riotto \cite{Kinney:2005in}. These arise from the ambiguity in the matching between present physical scales  and the number of e-folds before the end of inflation at which that scale left the horizon. Since we are  expanding the potential around some fiducial point, we never need to explicitly calculate the matching, and solve an exact relationship for the connection between $\ln{k}$ and the inflaton field $\phi$.  Furthermore, we can check that any given potential permits some minimum number of e-folds to elapse after cosmological scales have left the horizon.   On the other hand, any such calculation is likely to be inexact since the higher slow roll terms  can easily be significant at the {\em end\/} of inflation -- and inflation may end with a hybrid style transition induced by  an instability in a direction in field space that is orthogonal to the inflation.

The key assumption that underlies this work is that the primordial perturbations were generated by the standard inflationary mechanism -- and that during the epoch in which the perturbations are generated, the evolution of the universe can be accurately modeled in terms of a single, slowly rolling, minimally coupled scalar field.  We expect that if there is an accurate description of the early universe of this form, it will be via an effective theory, rather than one arising from the explicit  presence of an appropriate field in the Lagrangian that describes GUT scale physics.   In particular, there is a relation  between the overall excursion made by the inflaton field and  the energy scale of inflation \cite{Lyth:1996im,Easther:2006qu}. This can be used to argue, on the basis of effective field theory considerations, that it is very difficult for the inflaton to be a single, fundamental scalar field embedded within string theory.  However, models relying on multiple scalar fields do arise naturally in string theory (see e.g. \cite{Dimopoulos:2005ac,Easther:2005zr}), and the method described here will cover the effective single field description of these models, assuming that one exists.   Lastly, by using a truncated slow-roll hierarchy, we are implicitly assuming that the inflationary potential has no sharp ``features'' in the region traversed by the inflaton as observationally relevant scales leave  the horizon.  Models of this type were explicitly compared to CMB data in \cite{peiris/etal:2003}, and the computation of the perturbation spectrum in this situation frequently exceeds the capacity of the slow roll approximation \cite{Adams:2001vc}.  

This paper serves two purposes. On the one hand, it explores a new approach to extracting the most leverage from applying an inflationary prior to a cosmological dataset, while it also addresses the more topical question of the inflationary constraints that can be gleaned from the WMAPII dataset.    We present chains for the WMAPI+All dataset, which is comprised of the original WMAPI release \cite{Spergel:2003cb,Bennett:2003bz}, ACBAR \cite{Kuo:2002ua}, CBI \cite{Mason:2002tm}, VSA \cite{Grainge:2002da}, the HST prior on the Hubble constant \cite{Freedman:2000cf}, and a tophat age prior that  the universe is between 10 and 20 Gyr old.  We use the WMAPI+All chains to compare the different parameter estimation methodologies, and to verify our codes. We then apply the HSR reconstruction algorithm to the WMAP dataset \cite{Hinshaw:2006ia,Page:2006hz,Spergel:2006hy} on its own and to WMAPII+SDSS.\footnote{Sloan Digital Sky Survey \cite{Tegmark:2003uf}}  We will find that for these datasets,  the second order expressions for the running of the slow roll parameters (as in \cite{Leach:2002ar,Leach:2002dw,Leach:2003us})  yield essentially identical results to those obtained the full Hubble slow roll formalism. However, this work shows that the flow equations can be used directly in a Monte Carlo Markov Chain based parameter estimation and, as noted above, the HSR approach is ideally to the high quality, heterogeneous datasets we expect to be dealing with in the future.     Our specific results confirm that the combination of WMAPII and SDSS puts strong constraints on the inflationary parameter space, whereas WMAPII on its own is roughly equivalent to the WMAPI+All dataset. 

The rest of this paper is organized as follows. In the next section, we review the Hubble slow roll formalism and the computation of the perturbation spectra. In Section~\ref{sec:theory} we review the convergence issues and assumptions implicit in the use of a truncated slow roll hierarchy, in Section \ref{sec:mcmcanalysis} we describe the implementation of a Markov Chain Monte Carlo analysis with this parameter set, and give our results in Section \ref{sec:results} for both the WMAPI+All dataset, and the new WMAPII data \cite{Hinshaw:2006ia,Page:2006hz,Spergel:2006hy}.  We discuss our constraints on the slow roll parameters in Section \ref{sec:discussion}, and conclude in Section \ref{sec:conclude}.

\section{Slow Roll Hierarchy and the Inflationary Potential} \label{sec:flowrollmethod}

The inflationary flow hierarchy provides a convenient method for generating potentials satisfying slow roll conditions for use with Monte Carlo techniques. In this picture, one takes the view that the evolution of spacetime is governed by a single order parameter, which can be written in form of a minimally coupled scalar field. Under this assumption, one can reformulate the exact
dynamical equations for inflation as an infinite hierarchy of flow
equations described by the generalized ``Hubble Slow Roll'' (HSR)
parameters \cite{hoffman/turner:2001,kinney:2002,easther/kinney:2003,peiris/etal:2003,kinney/etal:2004}. In the
Hamilton-Jacobi formulation of inflationary dynamics, one expresses
the Hubble parameter directly as a function of the field $\phi$ rather
than a function of time, $H \equiv H(\phi)$, under the assumption that
$\phi$ is monotonic in time. The equations of motion for the
field and background are given by:
\begin{eqnarray}
\dot{\phi} &=&- \frac{\mpl^2}{4\pi} H'(\phi), \label{eq:phih}\\
\left[H'(\phi)\right]^2 -
\frac{12\pi}{\mpl^2}H^2(\phi)&=&-\frac{32\pi^2}{\mpl^4}V(\phi). \label{eq:hj}
\end{eqnarray}
Overdots correspond to time derivatives and primes denote
derivatives with respect to $\phi$. Equation \ref{eq:hj}, referred
to as the {\sl Hamilton-Jacobi Equation}, allows us to consider
inflation in terms of $H(\phi)$ rather than $V(\phi)$. The former,
being a geometric quantity, describes inflation more naturally.

The major advantage of this picture is that it removes the
field from the dynamics, and lets one study the generic
behavior of slow roll inflation without making detailed assumptions about the underlying particle physics.  The HSR parameters
$^{\ell}\lambda_H$ are defined by the infinite hierarchy of differential equations
\begin{eqnarray}
\epsilon(\phi) &\equiv& \frac{m^2_{\rm Pl}}{4\pi}
\left[\frac{H'(\phi)}{H(\phi)}\right]^2; \label{eq:eps} \\
^{\ell}\lambda_H &\equiv& \left(\frac{m^2_{\rm Pl}}{4\pi}\right)^\ell
  \frac{(H')^{\ell-1}}{H^\ell} \frac{d^{(\ell+1)} H}{d\phi^{(\ell+1)}}
   ;\ \ell \geq 1.  \label{eq:hier}
\end{eqnarray}
Substituting Eq.~\ref{eq:eps} into Eq.~\ref{eq:hj} immediately gives
the potential in terms of $\epsilon$:
\begin{equation}
H^2\left(\phi\right)\left[1-\frac{1}{3}\epsilon\left(\phi\right)\right] =
\left(\frac{8\pi}{3\mpl^2}\right)V(\phi). \label{eq:v}
\end{equation}

The flow equations allow us to consider the model space spanned by inflation. The trajectories of the flow parameters are governed by a set of coupled first order differential equations.  As first pointed out by Liddle \cite{liddle:2003}, these equations have an analytic solution. Truncating the hierarchy of flow parameters, so that the last non-zero term is $^M\lambda_H$ ensures that $^{M+1}\lambda_H = 0$ at all times. From Eq.~\ref{eq:hier}, it follows that
\begin{equation}
\frac{d^{(M+2)} H}{d\phi^{(M+2)}} = 0 
\end{equation}
at all times. We thus arrive at the general solution  $H(\phi)$ as an order  $M+1$ polynomial in $\phi$:
\begin{equation}
H(\phi) = H_0\left[ 1+ B_1 \left(\frac{\phi}{\mpl}\right) + \cdots + B_{M+1}
\left(\frac{\phi}{\mpl}\right)^{M+1}\right]. \label{eq:h}
\end{equation}
So far, we have not specified an initial value for $\phi$.   If we truncate the series at $^M\lambda_H$, any set of flow parameters spans an $M+1$ dimensional space. However, the evolution equations evolve the $^M\lambda_H$ backwards and forwards as a function of $\phi$. Consequently,  the set of distinct  trajectories spans an $M$ dimensional space, effectively fibering the space of initial conditions for the flow hierarchy. However, if the flow parameters are specified at $\phi=0$ the ambiguity is removed.    

Further, from the definition of $\epsilon(\phi)$, 
\begin{equation}
\epsilon(\phi) = \frac{m^2_{\rm Pl}}{4\pi}
\left[\frac{\left(\frac{B_1}{\mpl}\right) + \cdots + (M+1)
\left(\frac{B_{M+1}}{\mpl}\right)\left(\frac{\phi}{\mpl}\right)^M}{1+B_1\left(\frac{\phi}{\mpl}\right)
+ \cdots + B_{M+1}\left(\frac{\phi}{\mpl}\right)^{M+1}}\right]^2 \, .  \label{eq:epsanalytic}
\end{equation}
The coefficients $B_i$ can be written in terms of the initial values of the HSR parameters as
\begin{eqnarray}
B_1 &=& \sqrt{4\pi\epsilon_0} \\
B_{\ell+1} &=& \frac{(4\pi)^\ell }{(\ell+1)! 
\ B_1^{\ell-1}}  {}^{\ell}\lambda_{H,0};\ \ell \geq 1. \label{eq:coeffs}
\end{eqnarray}
The sign of $B_1$ specifies in which direction the field is
rolling. Without loss of generality, we can pick some fiducial
physical scale that corresponds to $\phi=0$ (which we choose to be
$k_0=0.002$ Mpc$^{-1}$). Then, with the above convention, $\phi>0$
corresponds to scales larger than $k_0$, and $\phi<0$ corresponds to
smaller scales. We  associate a physical wavenumber with a
value of $\phi$ by solving the additional differential equation,
\begin{equation}
\frac{d\phi}{d\ln k} = -\frac{\mpl}{2\sqrt{\pi}}
\frac{\sqrt{\epsilon}}{1-\epsilon}. \label{eq:phieq}
\end{equation}
Finally, we must express the scalar and tensor primordial power
spectra in terms of these parameters. A first order expansion around
an exact solution for the case of power law inflation
\cite{lidsey/etal:1995} gives
\begin{eqnarray}
P_{\cal{R}} &=&\left.  \frac{\left[1-(2C+1)\epsilon +
C\eta\right]^2}{\pi\epsilon}\left(\frac{H}{\mpl}\right)^2 \right|_{k=a H},
\label{eq:Pscalar} \\
P_h &=&\left. \left[1-(C+1)\epsilon\right]^2
\frac{16}{\pi}\left(\frac{H}{\mpl}\right)^2 \right|_{k=a H},
\label{eq:Ptensor}
\end{eqnarray}
where $\eta =\ ^{1}\lambda_H$, $C= -2 + \ln{2} + \gamma \approx -0.729637$ and $\gamma$ is the Euler-Mascheroni constant.   One minor inconsistency in our approach is that (\ref{eq:Pscalar}) and (\ref{eq:Ptensor}) are calculated using the slow roll approximation, and truncated at second order.  In practice,  we do not believe this is a significant drawback, as the datasets we are working with provide meaningful constraints on (at most) the lowest three slow-roll parameters.   In the future, this problem could be surmounted by either using a higher order expansion for the spectra, or even integrating the mode equations directly.

Now we have all the ingredients to use this formalism for
reconstructing the inflationary potential, using a Markov Chain Monte
Carlo (MCMC) analysis. Once we choose the order $M$ at which to
truncate the HSR series, we pick the values of the HSR parameters at
the fiducial scale, giving $H$ and  all the HSR parameters as  functions of $\phi$.  After solving for $k(\phi)$ we can map our results between $k$ and $\phi$,  giving us the primordial
power spectra as a function of $k$, from which we can then compute the CMB power spectra and estimate the cosmological parameters in the usual way. Finally, after constraining the HSR parameter values  via a MCMC analysis, Eq.~\ref{eq:v} can be used to reconstruct the corresponding family of potentials.  Finally, recall that $H(\phi)$ has an overall multiplicative normalization, $H_0$.   Within the slow roll hierarchy this is a free parameter.  However, once we specify the $B_i$ at $k_0$, $H_0$ is fixed by the amplitude of the observed scalar perturbation spectrum, $A_s$, after inverting equation \ref{eq:Pscalar}.\footnote{Note that the present day Hubble constant is specified by $h$, and is not connected with the $H_0$ used here.}

\section{Convergence Issues and Implicit Assumptions \label{sec:theory}}

It is worth noting is that we are effectively fitting to a specific functional form for the potential: once we specify $H(\phi)$ we have also specified $V(\phi)$.   To see this more clearly, let us cut off the slow roll expansion at $B_2$. In this case, 
\begin{eqnarray}
V(\phi) &=& \frac{\mpl^2 H_0^2}{32 \pi^2}\left[
  (12 \pi - B_1^2) + 4(6 \pi B_1 - B_1 B_2) \varphi + 4 (3 \pi B_1 + 6 \pi B_2 -4 B_2^2) \varphi^2 \right. \nonumber \\
     && \left. + 24 \pi B_1 B_2 \varphi^3 + 12 \pi B_2^2   \varphi^4 \right]
     \end{eqnarray}
where $\varphi = \phi/\mpl$.  This expression contains the $B_l$ at linear and quadratic order -- which reflects the $H^2$ and $H'^2$ terms rather than a slow roll related truncation -- including $B_3$ and higher terms in $H$ does not add terms of  the form $B_i B_j B_k $ to the potential.  Secondly, the overall scale of the potential is set by $H_0$, which fixes the energy scale of inflation.  Looking at the structure of $H^2$, we can see that the first $i$ powers of $\varphi$ will each have a linear contribution from $B_i$, up to $B_N$.   All other contributions involve quadratic combinations like $B_i B_j$ or $B_i^2$.  In particular, the $\varphi^{N+1}$ through $\varphi^{2N}$ terms cannot be fixed independently of the lower order terms. 

Near the fiducial point, $\varphi \ll 1$ and the higher order terms are effectively ignorable, and the only relevant contributions to $V(\phi)$ arise from the terms that are linear in $B_i$.  Thanks to a well-known result, originally due to Lyth \cite{Lyth:1996im,Easther:2006qu}, the change of $\phi$ during the course of inflation is a function of $r$, the  tensor:scalar ratio.  From our perspective, the two key results are
\begin{eqnarray}
r &=& 16 \epsilon \label{eq:rdef}  \, ,\\
\Delta \phi &=& \frac{\mpl}{8 \sqrt{\pi} } \sqrt{r} \Delta N \label{eq:Lythbound} \, .
\end{eqnarray}
The tensor:scalar ratio measures the relative amplitudes of $P_{\cal{R}}$ and $P_h$, while $\Delta N$ and $\Delta \phi$ measure the number of e-folds that elapse as the field rolls through $\Delta \phi$.\footnote{Note that several different definitions for $r$ can be found in the literature so one frequently encounters this relationship with a different numerical coefficient.}    These relationships are derived at lowest order in slow roll and can be extended \cite{Easther:2006qu} to higher orders, but for the present analysis these expressions are more than adequate. The immediate lesson to be drawn from equation~\ref{eq:Lythbound} is that if $r$ is well below the current bound, $\Delta \varphi$ will be less than unity even if $\Delta N$ is as large as 50.  

Recalling the relationship between $\epsilon$ and $B_1$ we can derive 
\begin{equation}
\frac{\Delta \phi}{\mpl}=  \frac{B_1}{4\pi} \Delta N \, ,
\end{equation}
again at lowest order in slow roll. Rewriting $H(\phi)$ in terms of the number of efolds that elapse since the $k_0$ mode crosses the horizon during inflation we find,
\begin{equation}
H(\Delta N) \approx H_0\left[  1+ B_1 \left(\frac{B_1 \Delta N}{4 \pi} \right) + \cdots + B_{M+1}
\left(\frac{B_1 \Delta N}{4 \pi}\right)^{M+1}\right]. \label{eq:hdeltan}
\end{equation}
Even with the constraint derived here from WMAPI+All data, we can confidently say that $\epsilon \le 0.02$ at $\phi_0$. If we are only interested in reconstructing of the piece of the potential that contributes to cosmologically relevant perturbations, we only need $B_i \lesssim 1$ for this expansion to be well behaved. However, if the bound on $\epsilon$ is saturated, then increasing $\Delta N$  sees $H(\Delta N)$  dominated by higher order terms and the approximation will become unreliable.   On the other hand, if it turns out that $\epsilon \ll 0.01$ then we could imagine pushing $\Delta N$ up to 60 without having to put tight constraints on the $B_i$ -- and in this case the end of inflation would presumably be encoded in  a divergence of the $B_i$ for large $i$.  This analysis is necessarily heuristic, but is allows us to establish a rough estimate for the region of validity of this approximation as a function of $\epsilon$.

The initial value of $\epsilon$ or $B_1$ effectively fixes $\Delta \phi$ during the interval in which observable modes leave the horizon. For the chains we run here, the constraints on $\epsilon$ from the WMAPI+All dataset guarantee that $\Delta \phi  < \mpl$ over the 8 e-folds following the moment when the $k_0$ mode crosses the horizon.  However, as observational bounds on $r$ tighten, the permissible range of $\Delta \phi$ will decrease.  Given that $B_2$ is typically very much less than unity, the expansion appears to be well behaved in the region of interest, but this expansion is almost certainly asymptoically rather than absolutely convergent, and would eventually diverge if a sufficiently large number of terms was included.

\section{Markov Chain Monte Carlo Analysis} \label{sec:mcmcanalysis}

We use a Markov Chain Monte Carlo (MCMC) technique \cite{Christensen:2000ji, Christensen:2001gj,Knox:2001fz,Lewis:2002ah,Kosowsky:2002zt,Verde:2003ey} to evaluate the likelihood function of model parameters. The MCMC is used to simulate observations from the posterior distribution ${\cal P}({\bf \alpha}|x)$, of a set of parameters ${\bf \alpha}$ given event $x$, obtained via Bayes' Theorem,
\begin{equation}
{\cal P}(\alpha|x)=\frac{{\cal P}(x|\alpha){\cal P}(\alpha)}{\int
{\cal P}(x|\alpha){\cal P}(\alpha)d\alpha},
\label{eq:bayes}
\end{equation}
\noindent where ${\cal P}(x|\alpha)$ is the likelihood of event $x$ given the model parameters $\alpha$ and ${\cal P}(\alpha)$ is the prior probability density. The MCMC generates random draws (i.e. simulations) from the posterior distribution that are a ``fair'' sample of the likelihood surface. From this sample, we can estimate all of the quantities of interest about the posterior distribution (mean, variance, confidence levels). A properly derived and implemented MCMC draws from the joint posterior density ${\cal P}(\alpha|x)$ once it has converged to the stationary distribution. 

We use modified versions of the CosmoMC\footnote{{\tt http://cosmologist.info/cosmomc/}}/CAMB\footnote{{\tt http://camb.info/}} packages for our computations, utilizing eight chains per model and a conservative convergence criterion \cite{gelman/rubin:1992}. For our application, $\alpha$ denotes a set of cosmological parameters, the number of which varies with the model chosen to describe the primordial power spectrum (see below). We always vary 4 ``late-time'' parameters: the physical energy density in baryons $\Omega_b h^2$, the total physical energy density in matter $\Omega_{CDM} h^2$, the optical depth to reionization $\tau$, the Hubble constant $h$ in units of $100$ km s$^{-1}$ Mpc$^{-1}$.   The ``event'' $x$ is the  likelihood of obtaining the total observational dataset (either WMAPI+All, WMAPII or WMAPII+SDSS in this paper)  if we assume that the actual universe has an underlying set of parameter values given by $\alpha$.

In order to understand the distinctions between the Hubble slow roll parameters and the characterization of the spectrum in terms of an index and running, we perform three distinct analyses.   Firstly, we present results for the traditional power law primordial observables: the scalar spectral index, $n_s$, its running with scale, $dn_s/d\ln k$, the tensor to scalar ratio, $r$, the tensor spectral index, $n_t$, and the scalar spectral index, $A_s$ (all defined at the fiducial scale $k_0=0.002$ Mpc$^{-1}$)  We are including the amplitude and indices for both the scalar and tensor spectra. In single field slow roll inflation these values are related.  However, we wish to avoid the use of any inflationary priors in these chains, but many analyses adopt this ansatz in practice since the tensors are currently unobserved.  So far as we know, this particular parameter set has not be used before in this context -- as one would expect, parameters corresponding to  the (currently unobserved) tensor power spectra are very weakly constrained.  We denote this formulation by $\{n_s,n_t\}$.

Secondly, we run chains where the  HSR parameters are varied inside of the Markov chains, but the spectra are computed by converting these values into ``traditional'' power law variables, using the formulae below:
\begin{eqnarray}
n_s &=& 1+ 2\eta - 4\epsilon - 2(1+{\cal C}) \epsilon^2 - \frac{1}{2}(3-5{\cal C}) \epsilon \eta + \frac{1}{2}(3-{\cal C})\xi \\  \label{eq:hsrconversionsstart}
r &=& 16 \epsilon \left[1+2{\cal C}(\epsilon - \eta)\right]\\
n_t &=& -2\epsilon - (3+{\cal C}) \epsilon^2 + (1+{\cal C}) \epsilon \eta \\
\frac{dn_s}{d\ln k} &=& -\frac{1}{1-\epsilon} \left\{2 \frac{d\eta}{dN} - 4 \frac{d\epsilon}{dN} - \right.  \nonumber \\
&& \quad \left. 4\left(1+{\cal C}\right) \epsilon\frac{d\epsilon}{dN} - \frac{1}{2}\left(3-5{\cal C}\right)\left(\epsilon\frac{d\eta}{dN} + \eta \frac{d\epsilon}{dN}\right) + \frac{1}{2}\left(3-{\cal C}\right)\frac{d\xi}{dN} \right\}
\end{eqnarray}
where ${\cal C}=4(\ln 2 + \gamma) - 5$, and we note that ${\cal C}$ differs from the $C$ in the expressions for the spectra.  We refer to this formulation as $n_s(\epsilon,\eta)$.

The scale dependence of the slow roll parameters is obtained directly from the flow equations, 
\begin{eqnarray}
\frac{d\epsilon}{dN} &=& 2\epsilon(\eta-\epsilon)\\
\frac{d\eta}{dN} &=& -\epsilon\eta + \xi \\
\frac{d\xi}{dN} &=& \xi (\eta-2\epsilon) +\  ^3\lambda_H \label{eq:hsrconversionsend}
\end{eqnarray}
which we have truncated at third order by setting $^3\lambda_H =0$. In practice we will set $\xi = \ ^2\lambda_H = 0$ as well, so that only $\eta$ and $\epsilon$ vary as the universe evolves.  Finally, $N$ denotes the number of e-folds of inflation, which does not not explicitly appear in this formulation. 

The initial values of the HSR parameters are assumed to be set at the fiducial scale $k_0=0.002$ Mpc$^{-1}$. The third and final parameterization uses the HSR parameters directly, calculating the power spectrum using the formulae \ref{eq:Pscalar} and \ref{eq:Ptensor}. The power spectrum is normalized at $k_0$ by setting
\begin{equation}
A_s = \frac{\left[1-(2C+1)\epsilon_0 +
C\eta_0\right]^2}{\pi\epsilon_0}\left(\frac{H_0}{\mpl}\right)^2, \label{eq:norm}
\end{equation}
where $\epsilon_0$ and $\eta_0$ are the values at $k_0$.

\TABLE[!htb]{
\begin{tabular}{||c|c|c|c||}
\hline
  & Dataset & Variables & Comments \\
\hline 
1. & WMAPI+All &$ \{n_s,n_t, \cdots\}$ &  Standard spectral variables \\
\hline  
2. & WMAPI+All &$ \{n_s(\epsilon,\eta),n_t(\epsilon,\eta), \cdots\}$ & 
Slow roll parameters in chains  \\
\hline
3. &  WMAPI+All & HSR $\{\epsilon,\eta\}$ & 2D HSR \\
\hline
4. &WMAPII & HSR $\{\epsilon,\eta \}$ & 2D HSR \\
\hline
5. &WMAPII + SDSS & HSR $\{\epsilon,\eta \}$ & 2D HSR \\
\hline
\end{tabular}
\caption{ MCMC Parameter fits  described in this paper.   }\label{table:summary}}

\section{Results  \label{sec:results}}

We now turn to the results of our experimentation with different characterizations of the primordial spectrum.    In all the chains discussed below, we work with the following set of ``late time'' parameters, $\Omega_b h^2$, $\Omega_{CDM} h^2$, $h$ and $\tau$, where these variables have their usual meanings.  Without careful tuning, the inflationary universe is extremely close to flat, so it is consistent with our imposition of an inflationary prior to set $\Omega_{\mbox{\tiny Total}} = 1$. The contribution from dark energy is thus $\Omega_\Lambda = 1- \Omega_b - \Omega_{CDM}$, and we take its equation of state to be $w_\Lambda=-1$. We assume instantaneous reionization at an optical depth of $\tau$, the HST Key Project flat prior  on the Hubble parameter \cite{Freedman:2000cf}, and a flat prior on the age of the universe, as described in the Introduction.

In the 1D plots that follow,  we display the probability distributions for
each parameter marginalized over all the other parameters. In the 2D 
figures, the solid contours are the marginalized 1 and 2 $\sigma$ joint 
probability contours for each pair of parameters.   Each plot contains data drawn from 8 chains, which were well-converged according to the Gelman-Rubin criterion.  The CMB spectra were computed with an appropriately modified version of CAMB, and the chains themselves run with the CosmoMC package.

We ran Monte Carlo Markov Chains over the WMAPI+All dataset for a three different parametrizations of the fundamental spectral variables, as outlined in Table~1.   In Model 1, the primordial parameters $\{n_s, dn_s/d\ln k, r, n_t, A_s\}$ measured at $k_0=0.002$ Mpc$^{-1}$ were varied within the chains.   This covers much the same ground as many previous analyses  (e.g. \cite{Bennett:2003bz}) -- the one novelty being that we have fitted to both the amplitude and index of the tensor spectrum. As might be expected in the absence of an observed tensor contribution to the primordial perturbations, these parameters are only  very weakly constrained.   The one constraint we do make is that $n_t< 0$, so that the tensor spectrum is red.\footnote{The height of the tensor spectrum scales with the energy density of the universe -- and this is strictly decreasing unless one has either ``phantom'' energy with $w<-1$ \cite{Caldwell:1999ew}, or if the perturbations are laid down during a contracting phase, as in the ekpyrotic  \cite{Khoury:2001wf}
 and pre-big bang  \cite{Gasperini:1992em} scenarios. These models do have blue gravitational spectra, but the tensor amplitude at cosmological scales is typically infinitesimal.} However, such models also typically have a vanishingly small tensor amplitude at CMB scales, so it is a very mild assumption.  We then ran fits using the two different slow roll schemes -- that of Leach {\em et al.\/} \cite{Leach:2002ar,Leach:2002dw,Leach:2003us}, and the HSR formalism described in this paper. We see that up to second order they give very similar results, as we would have expected.   Finally, we repeat the process for WMAPII on its own, and WMAPII+SDSS, obtaining constraints on $\epsilon$ and $\eta$ via the HSR formalism.

\FIGURE[!ht]{
\epsfig{file=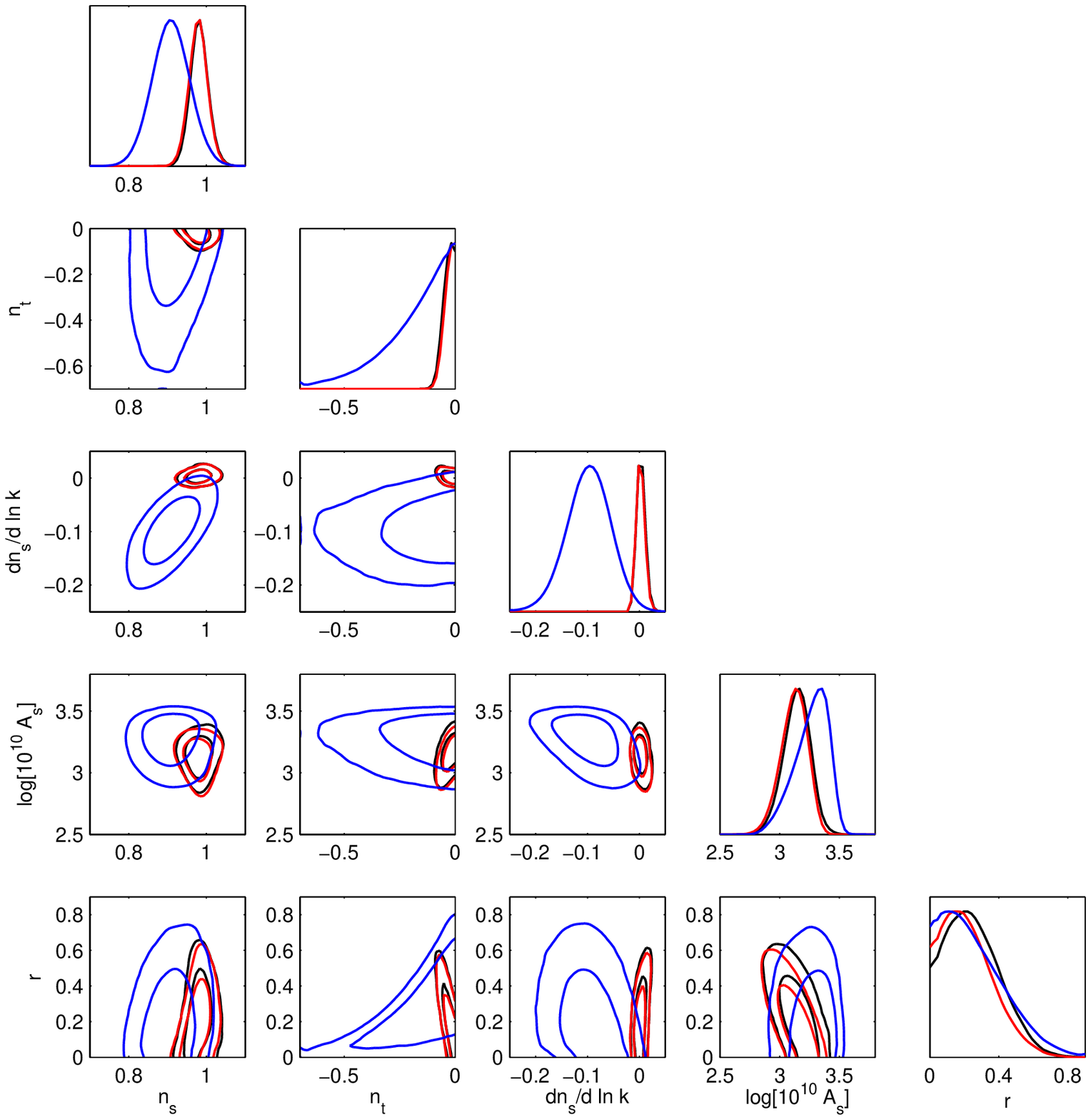,scale=0.85}
\caption{The results of three different fits to the WMAPI+All dataset. The blue lines are derived from the usual scalar and tensor spectral indices (Model 1). The black lines are the results of a fit using the $n_s(\epsilon,\eta)$ approach (Model 2), and the red lines correspond to the HSR algorithm described in this paper (Model 3). The top plot in each column shows the probability distribution function for each of the different ``primordial'' variables, which the other plots show their joint 68\% and 95\% confidence levels.   When we impose an inflationary prior, we convert $\epsilon$ and $\eta$ back into the parameter values at $k_0$.}\label{fig:pl}}

\subsection{WMAPI+All with an Inflationary Prior}

\FIGURE[!ht]{
\epsfig{file=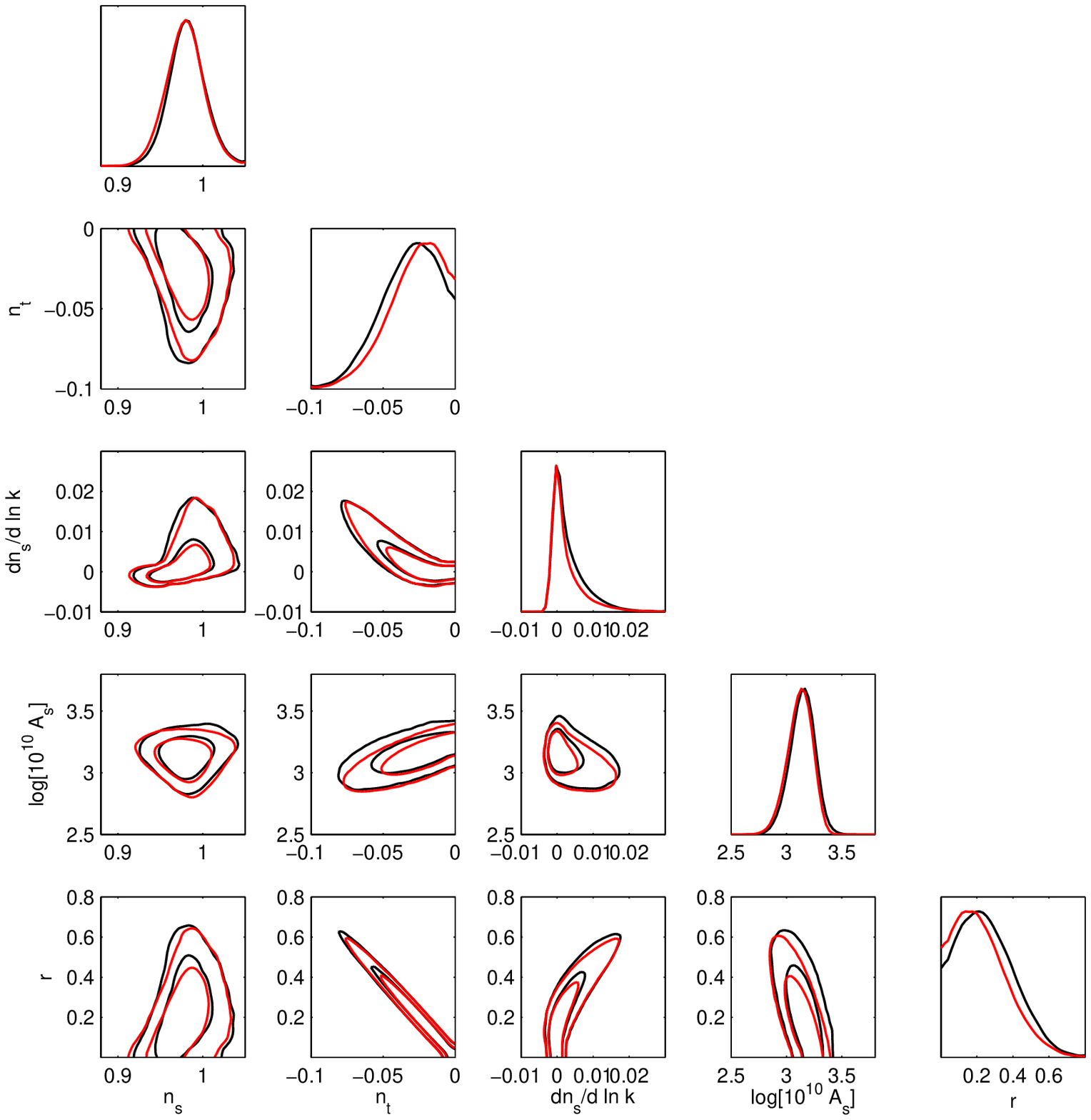,scale=0.85}
\caption{The two different slow roll fits to the WMAPI+All dataset, with the same conventions as Figure~\ref{fig:pl}. We see that the two sets of contours are essentially identical. }\label{fig:pl2}}

\FIGURE[!ht]{
\epsfig{file=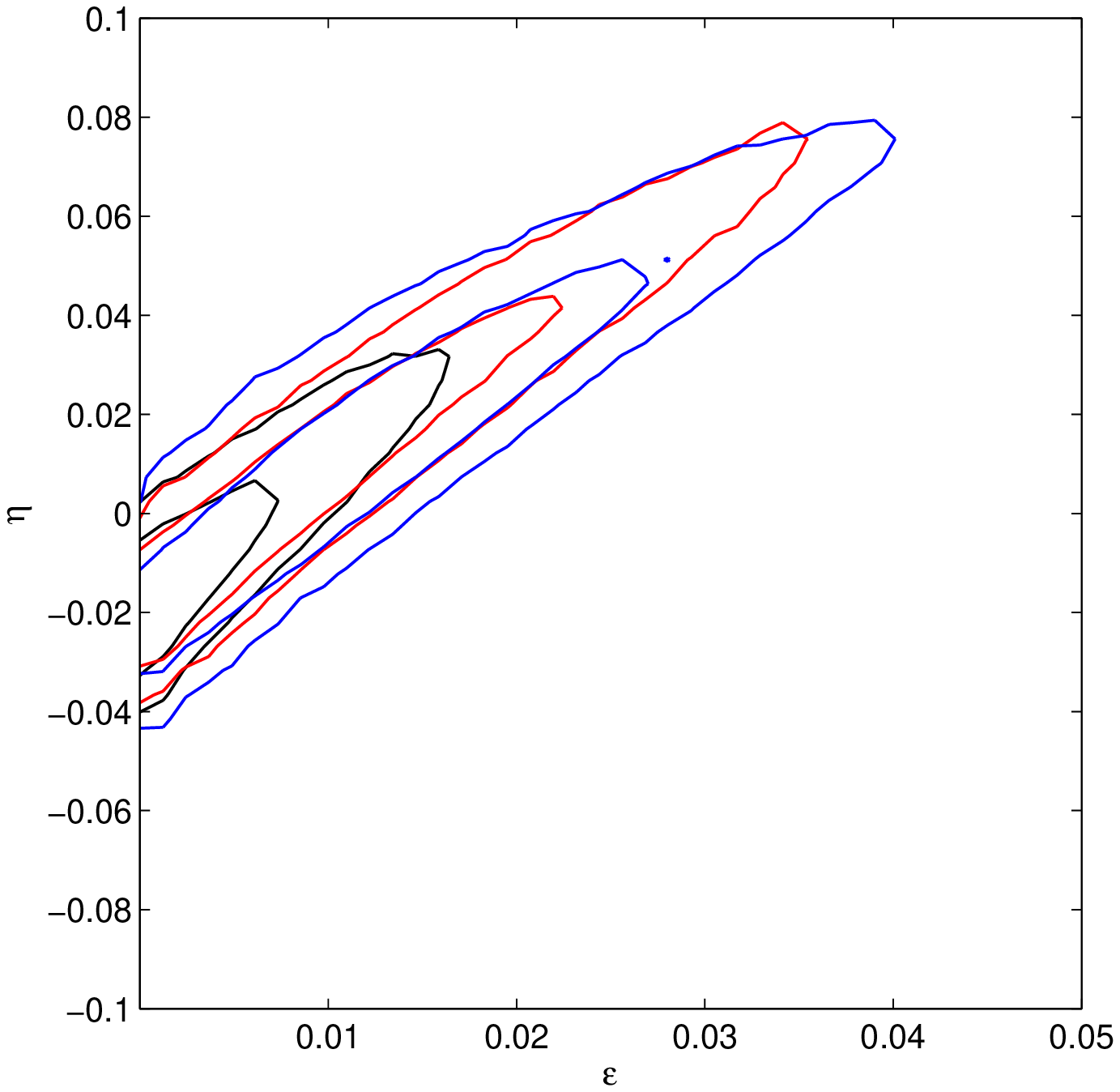,scale=0.85}
\caption{The allowed values of $\epsilon$ and $\eta$ derived from the MCMC chains for Models  3, 4 and 5. In each case we show the  joint $68\%$ and $95\%$ confidence contours.  The blue lines are WMAPI+All (Model 3), the red lines are WMAPII (Model 4) and the black lines are WMAPII+SDSS (Model~5)   }  \label{fig:pl3}}

Looking at Figure \ref{fig:pl}, we see that the with the inclusion of an arbitrary tensor spectrum and in the absence of any inflationary prior, the allowed parameter ranges are very broad when compared to the values typically extracted from fits to recent datasets. We can attribute this to the contribution of the tensor component at low $\ell$, which can be very large in these fits.  Moreover,  we see that a small value of $r$ is correlated with a very red tensor spectrum. In this case the tensor modes contribute strongly at the longest scales, and the normalization has to be reduced in order to obtain the correct normalization for the lowest lying $C_\ell$'s.   Likewise, we see evidence (at the 2$\sigma$ level) for a substantial running in the scalar index. This mirrors the result for $\alpha$ obtained from the WMAPII dataset \cite{Spergel:2006hy}.  

Conversely, when we express the spectrum in terms of $\epsilon$ and $\eta$ we see that the available region of parameter space is dramatically reduced - particularly in the case of the tensor modes, and $dn_s/d\ln{k}$, which reflects the imposition of the slow roll prior.    In particular, many analyses of the primordial  perturbations impose a slow roll prior that relates the amplitude and slope of the tensor perturbation spectrum, and doing so dramatically reduces the range of the $(n_t,r)$ plane  that is compatible with the data.   Likewise, the natural ``center of gravity'' for the WMAPI+All dataset prefers a negative value of $\alpha$, the running of the scalar spectral index.  Provided we use only the first two terms in the slow roll expansion ($\epsilon$ and $\eta$), this quantity can be large only if the scalar spectral index is far from unity.  If the central values of $\alpha$ found  from these two different parametrization do not change in future datasets, but the error ellipses contract, then this would provide an alternative ``consistency'' check on simple forms of slow roll inflation. Any failure of the data to satisfy would demonstrate either than the underlying inflationary model has a non-trivial potential, or that the perturbations are inconsistent with the assumption that they are generated by an inflationary epoch generated by a single, minimally coupled scalar field.   We fit the WMAPI+All dataset using both Leach {\em et al.\/}'s approach (Model 2) and our HSR algorithm (Model 3). The parameter values obtained for these fits are shown on their own in Figure \ref{fig:pl2}.  The two approaches lead to very similar results for the ``early universe'' cosmological parameters. In Figure \ref{fig:pl3} we show the constraints on $\epsilon$ and $\eta$ for the HSR fit. The most obvious feature here is the degeneracy direction following from  $n_s -1 \approx 2\eta - 4\epsilon $.

\FIGURE[!ht]{
\epsfig{file=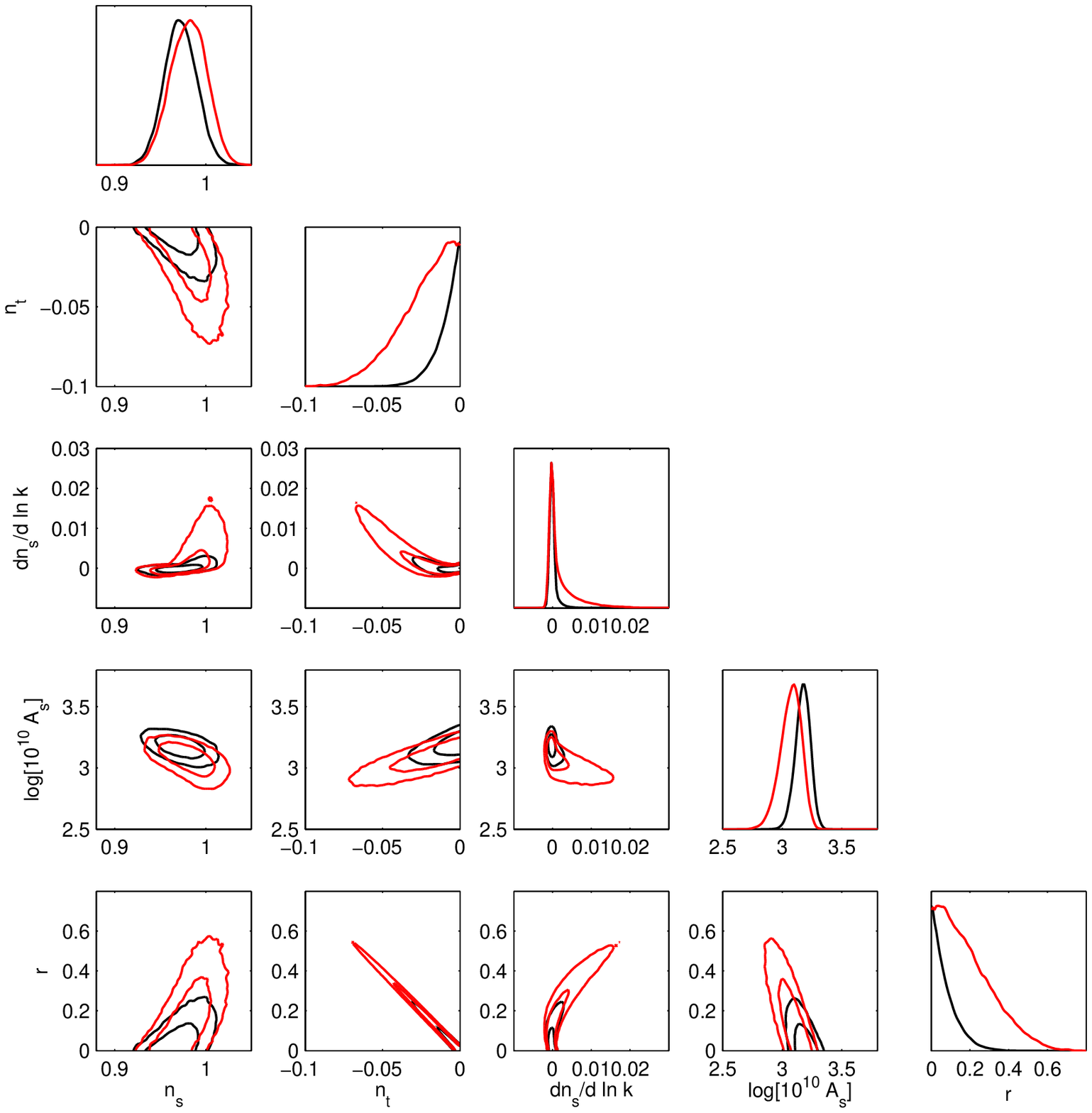,scale=0.85}
\caption{We show the results for using the HSR formalism to apply an inflationary prior to  WMAPII (red) and WMAPII+SDSS (black). We again show joint $68\%$ and $95\%$ confidence contours. \label{fig:pl4}}}

\TABLE[!htb]{
\begin{tabular}{||c|c|c||}
\hline
Parameter           &  WMAPII              &  WMAPII+SDSS \\
\hline 
$n_s$               & $0.982 \pm 0.020$    &  $0.973 \pm 0.017$ \\
\hline 
$d n_s/d\ln k$      & $0.002^{+0.003}_{-0.004}$  &  $0.0000^{+0.0002}_{-0.0005}$ \\
\hline
$\ln[10^{10} A_s]$ & $3.07 \pm 0.09$      &  $3.17 \pm 0.06$ \\
\hline
$r$                 & $<0.465$ (95\% CL)   &  $<0.200$ (95\% CL) \\
\hline
$n_t$               & $>-0.059$ (95\% CL)  &  $>-0.025$ (95\% CL) \\
\hline
\end{tabular}
\caption{ Constraints on primordial parameters, 
defined at $k=0.002$ Mpc$^{-1}$.  Constraints are at the 68\% level unless otherwise noted.}\label{table:constraints}}

\subsection{WMAPII and Inflation}

We use the HSR formalism to apply an inflationary prior to the WMAPII (Model 4) and WMAPII+SDSS (Model 5) datasets.  The constraint on $\epsilon$ and $\eta$ is shown in Figure \ref{fig:pl3}, and we immediately see that WMAPII on its own constrains the data slightly more strongly than the previous ``global'' dataset that comprises WMAPI+All.  However, the more interesting result here is WMAPII+SDSS, which we  see leads to a considerably tighter set of constraints, and starts to break the degeneracy between $\epsilon$ and $\eta$ seen in all the other datasets.  

We plot the corresponding parameter values in Figure \ref{fig:pl4}, and present the corresponding numerical bounds in Table~\ref{table:constraints}.  We immediately see that there are strong constraints on the tensor contribution and the running of the scalar index for the WMAPII+SDSS case.  In both cases these arise via the breaking of the degeneracy that allows $\epsilon$ to run out to relatively large values.  The joint 95\% confidence interval for the value of $r$ is lower than is obtained for a generic $(n_s,r)$ fit on the same dataset \cite{Spergel:2006hy}.  Moreover, the central value of $n_s$ has moved slightly towards unity, relative to \cite{Spergel:2006hy}, though the distributions are consistent at less than the $1\sigma$ level. Consequently, while the the pure Harrison-Zel'dovich spectrum is disfavoured, it is not strongly excluded.  Finally, we note that this data continues to be a good for fit for $m^2 \phi^2$ inflation, and $\lambda \phi^4$ is excluded at more than $2\sigma$-joint by the WMAPII+SDSS fit on account of its primordial tensor spectrum alone.

 Note that for the WMAPII data, we use an updated beam error module which differs from the version that was used in \cite{Spergel:2006hy}. This is available as an option in the public WMAPII likelihood software,\footnote{However, it is turned off by default.} and uses the full off-diagonal $C_l^{TT}$ covariance matrix, a Gaussian+$\ln$ normal approximation for the $C_l^{TT}$ likelihood, and a fixed fiducial $C_l^{theory}$ to propagate the beam errors in the likelihood function. The default options in the likelihood software assume a diagonal off-diagonal $C_l^{TT}$ covariance matrix, a Gaussian approximation for the $C_l^{TT}$ likelihood, and $C_l^{theory}$ varied in the steps in the MCMC to propagate the beam error. Note that, unlike in \cite{Spergel:2006hy}, we do not marginalize over the amplitude of the SZ fluctuations. This should be a mild assumption since there is no detection of SZ fluctuations, and the expected amplitude of SZ contamination in the data is at the level of the weak lensing of the CMB which was also neglected. Also, adding such a parameter could cause spurious degeneracies within our parameter set, a consequence which we wanted to avoid.  
 
We show our constraints from WMAPII and the closest corresponding constraints from \cite{Spergel:2006hy} (a fit using $r$ and $n_s$ only, with no running) in Figure \ref{fig:spergel}. The differences in the likelihood approximation between our analysis and that of Spergel {\em et al.\/} outlined above only shifts the constraints by considerably less than $1\sigma$-marginalized if the $(r, n_s)$ parametrization is used with the differing beam treatments and the SZ marginalization. Therefore the differences in the  distributions for $n_s$ and $r$ seen in Fig 5 arise from the imposition of the slow roll  prior.
 
\FIGURE[!ht]{
\epsfig{file=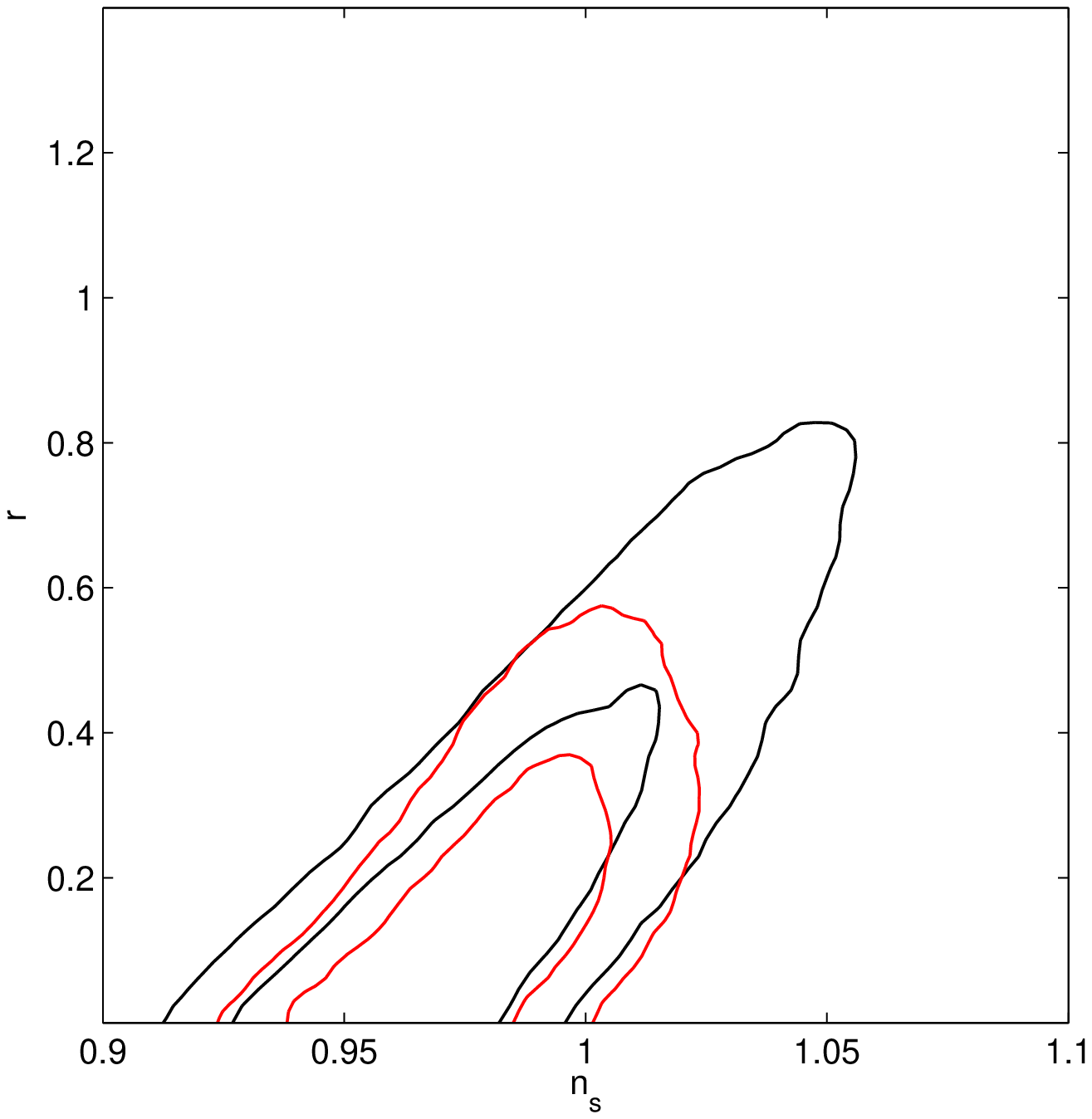,scale=0.85}
\caption{Comparison of constraints in the $r$ vs. $n_s$ plane: The red lines correspond to the HSR algorithm for WMAPII described in this paper (Model 3)  converting $\epsilon$ and $\eta$ back into the plotted parameter values at $k_0$. The black lines are for a fit to the power law variables $r$ and $n_s$  for the same data set. In both cases, joint 68\% and 95\% confidence levels are plotted.}\label{fig:spergel}}

\section{Discussion \label{sec:discussion}}

\FIGURE[!ht]{
\epsfig{file=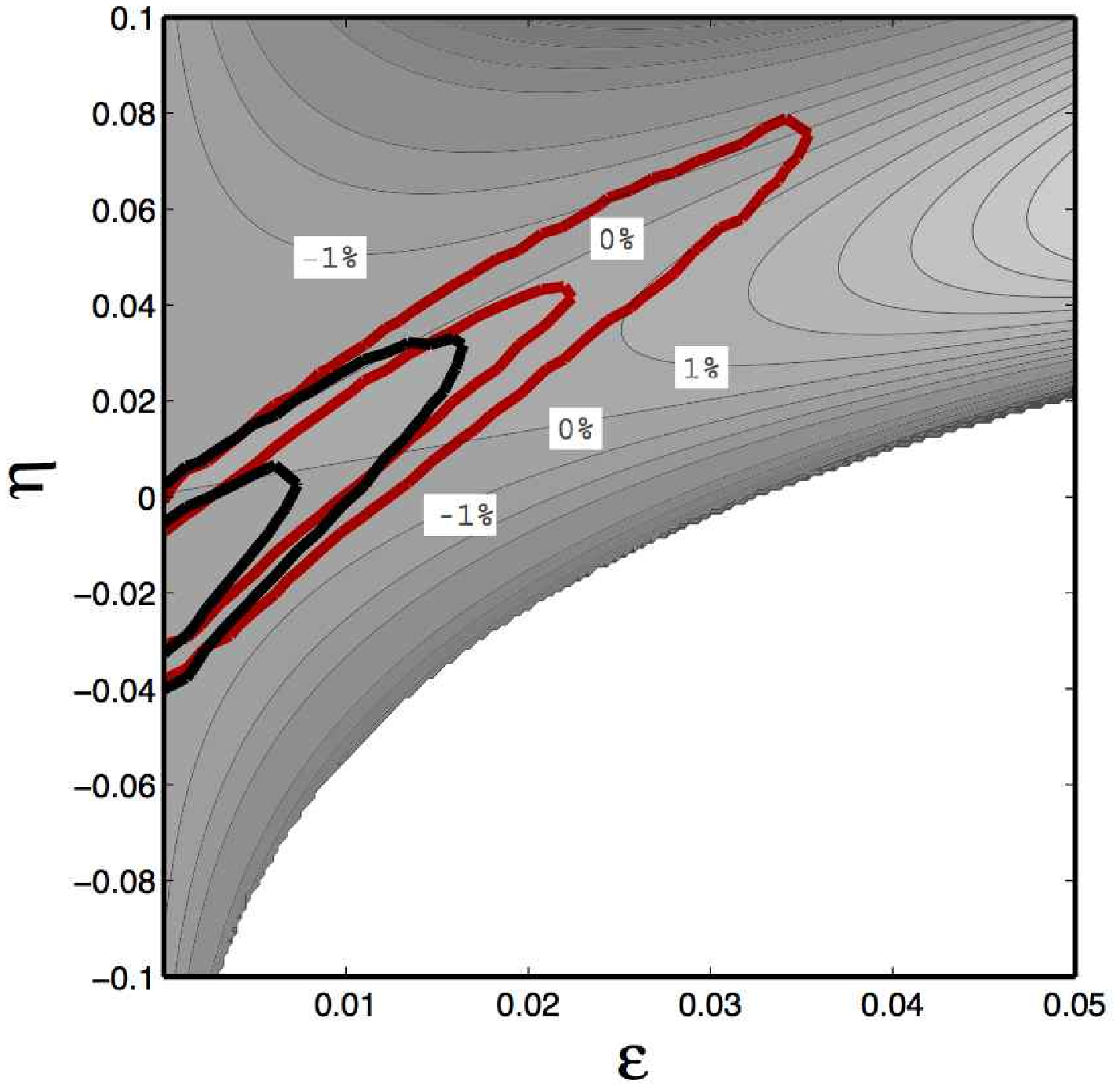,scale=1.0}
\caption{  We plot the difference between the scalar power spectrum computed using $n_s(\eta, \epsilon)$ and the full HSR parameters at $1$ Mpc$^{-1}$. The contours are $1\%$ apart, and a positive difference corresponds to the HSR result being larger than the $n_s(\epsilon,\eta)$ value. We have overlapped the joint 68\% and 95\% confidence intervals for $\epsilon$ and $\eta$ obtained for WMAPII (red) and WMAPII+SDSS (black). The white region corresponds to parameter values which support less than 15 e-folds of inflation after the $k_0$ mode leaves the horizon. \label{fig:pl5} }}

\FIGURE[!ht]{
\epsfig{file=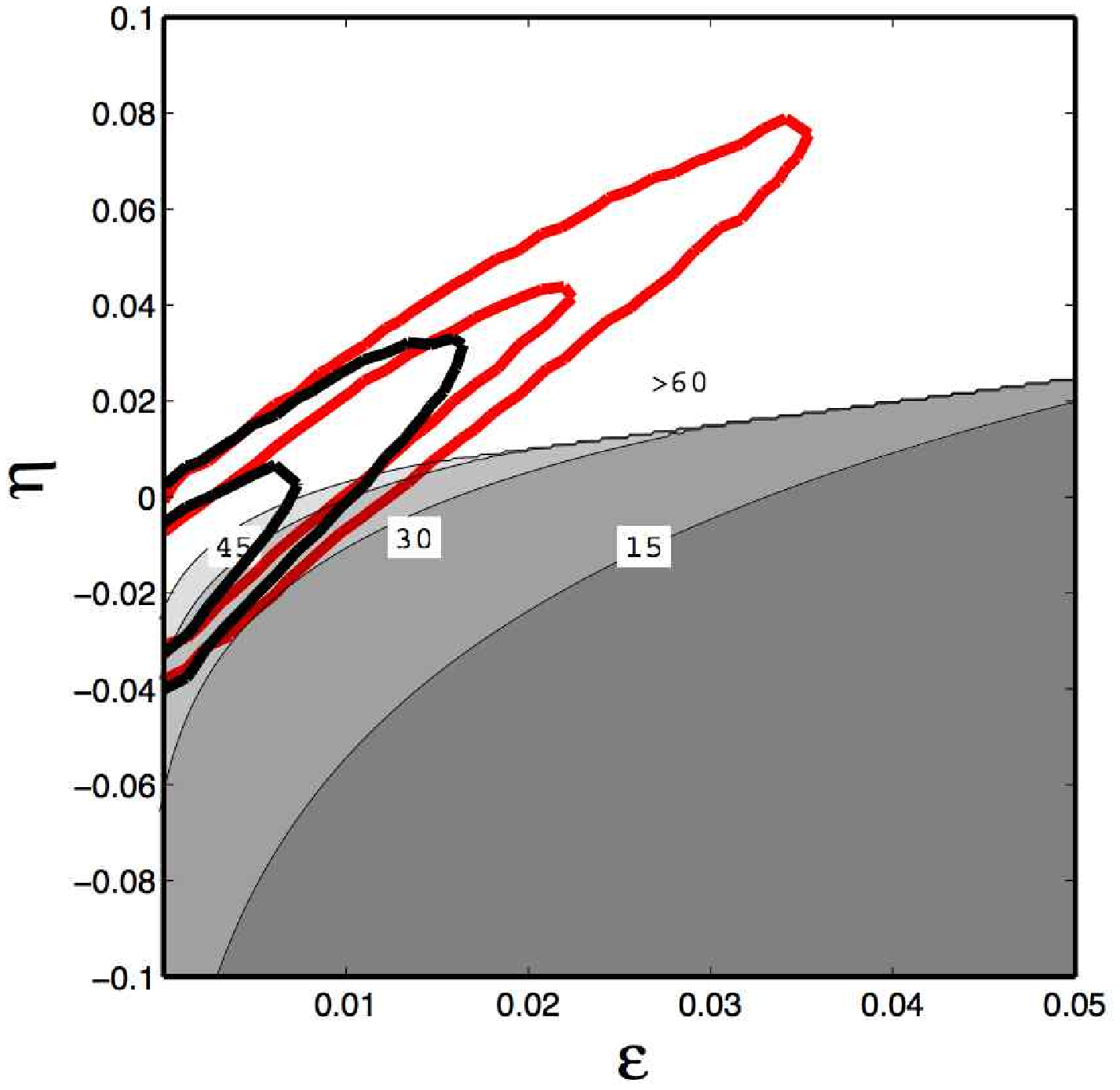,scale=1.0}
\caption{  We show the number of e-folds that elapses between the slow roll equations becoming singular (marking the end of inflation) and the mode $k_0$ leaving the horizon, along with the WMAPII and WMAPII+SDSS constraints, with the same color coding as Figure \ref{fig:pl5}. }\label{fig:pl6}
}

\FIGURE[!ht]{
\epsfig{file=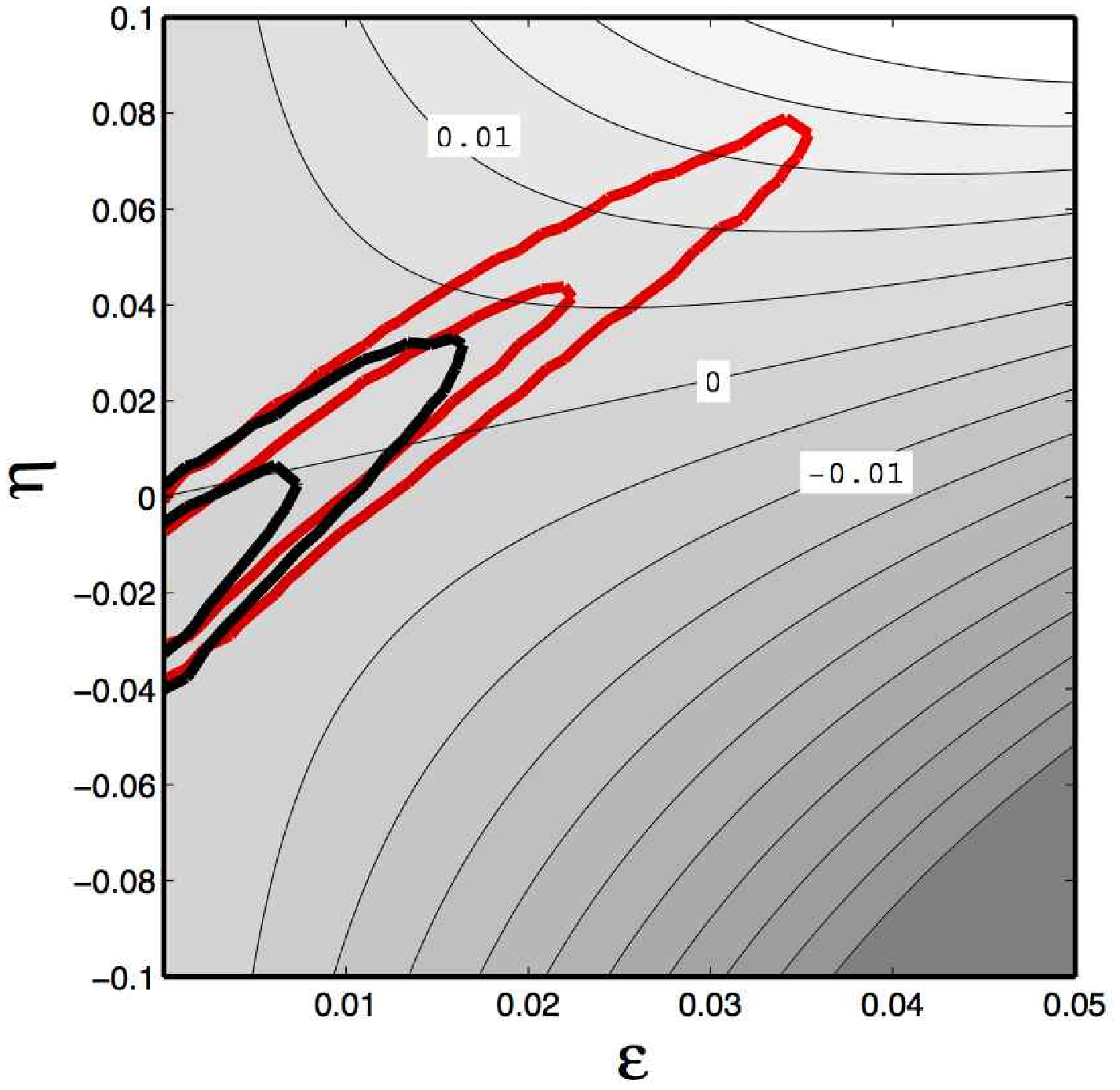,scale=1.0}
\caption{$d n_s /d \ln{k}$ is shown as a function of $\epsilon$ and $\eta$, along with our constraints on $\epsilon$ and $\eta$.  }\label{fig:pl7}}

\FIGURE[!ht]{
\epsfig{file=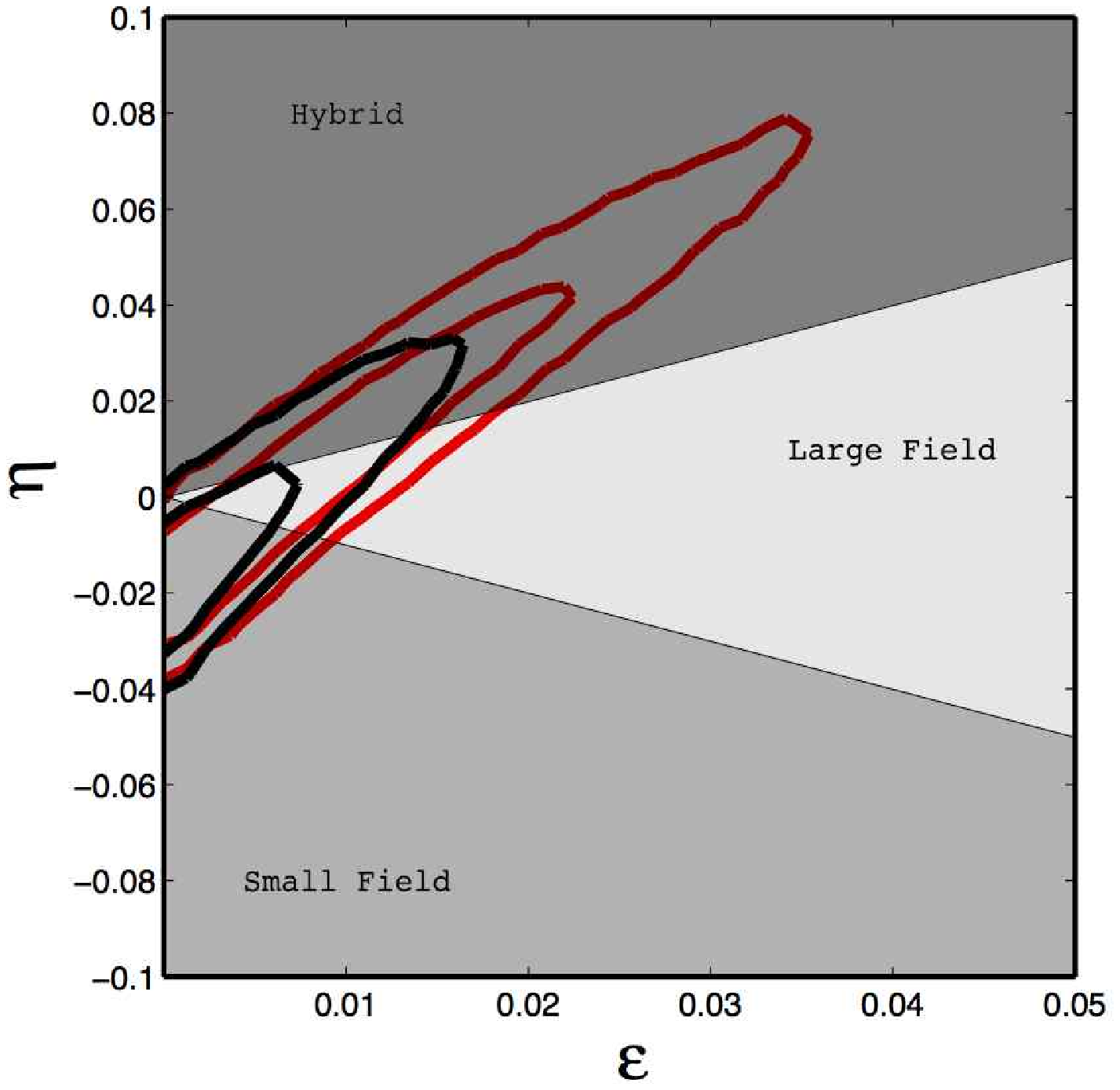,scale=1.0}
\caption{The inflationary parameter splace is shown as a function of $\epsilon$ and $\eta$, along with our constraints on $\epsilon$ and $\eta$.  }\label{fig:pl8}}

The results in the previous section display the effectiveness of the flow reconstruction algorithm, and clarify the distinctions between this approach and other methods that have been tried in the past.  To recapitulate, the primary advantages of this approach is that is does not require one to  Taylor expand the perturbation spectrum about an arbitrary fiducial point, and working with the full inflationary dynamics  is guaranteed to capture any consistency relationships that relate the spectral parameters.   In particular, this process represents a step forward from Monte Carlo reconstruction, in that it avoids the use of an arbitrary ``measure'' on the parameter space occupied by the initial values of the flow parameters. 

As stated in the Introduction, a similar approach to the problem of determining the values of the slow roll parameters (and from them, the inflationary potential) has been proposed by Leach and collaborators  \cite{Leach:2002ar,Leach:2002dw,Leach:2003us}.  Their method is based on Taylor expanding the slow roll parameters about their values at some fiducial point.  For many purposes this method is equivalent to the flow equation approach developed here, but it differs in two crucial ways. Firstly, the flow equations allow reconstruction to be carried out to arbitrary order in the slow roll parameters, and secondly  when we use the flow equations we can explicitly check that our choice of slow roll parameters is consistent with obtaining the required amount of inflation. In particular, unless $\epsilon$ is vanishingly small, we typically require at least 50 efolds of inflationary growth to occur after the mode $k_0 = 0.002 \mbox{Mpc}^{-1}$  leaves the horizon.  Looking at Figure \ref{fig:pl6} we see that imposing this requirement actually removes a piece of the parameter space that overlaps with the allowed regions for $\epsilon$ and $\eta$ that one obtains from the MCMC fits.  We caution against simply imposing an additional constraint on the parameter space by insisting that $N>55$ -- however, we can say that models in this apparently excluded region must have a significant contribution from higher order terms in the slow roll expansion in order to be viable.  

More generally,  we have chosen to expand $H(\phi)$ as a polynomial in $\phi$. While this is perhaps a natural choice it is certainly not unique -- in principal one could use any set of non-degenerate functions of $\phi$ as a basis for the MCMC fit.   In particular, if one wanted to generalize this method to potentials with ``features'' it may be useful to introduce a different parameterization of $H(\phi)$.\footnote{In this context see also the analysis of \cite{Kadota:2005hv}.}  In this case we may need to replace Eqs.~\ref{eq:Pscalar} and \ref{eq:Ptensor}  with a direct integration of the differential equations describing the evolution of the scalar and tensor perturbations to calculate their power spectra as a function of $k$, before feeding them into the CMB power spectrum calculation.

In Figures \ref{fig:pl5} through \ref{fig:pl8} we explore our different constraints on $\epsilon$ and $\eta$.\footnote{In all plots, we are showing $\epsilon_0$ and $\eta_0$, the instantaneous values of the slow roll parameters when $k=k_0$, but we drop the $0$ subscripts for clarity.}   Figure \ref{fig:pl5}  shows the difference  in the primordial power spectra  (not the $C_\ell$'s themseves) computed directly from the HSR expansion, and the values obtained using the $n_s(\epsilon,\eta)$ approach of Leach {\em et al}. We see that these overlap well, except for relatively extreme values  of the slow roll parameters.   However, as we discuss in the Introduction, the use of heterogeneous datasets may eventually lead to constraints on the primordial power spectrum for modes with $k$ considerably smaller than 1 Mpc$^{-1}$, and in this case the HSR formalism will be indispensable.  We plan to address this issue in more detail in future work.

In Figure \ref{fig:pl6} we show the number of e-folds that elapse after the mode $k_0$ leaves the  horizon, as a function of $\epsilon$  and $\eta$.\footnote{Strictly speaking, we plot the number of e-folds that elapses before the slow roll hierarchy becomes singular, with $\epsilon\rightarrow \infty$. In practice this is effectively synonymous with the end of inflation. The singularity has no physical significance, since the HSR expansion breaks down when inflation ends.} This result has to be interpreted with care, since while higher order terms in the slow roll expansion may have a negligible  impact on the spectrum at CMB scales, they can exert considerable leverage between the generation of cosmological perturbations and the end of inflation.  However, this shows it is straightforward to add constraints on the total number of e-folds of inflation directly to the parameter estimation process.  Moreover, we see that a naive imposition of the requirement that $N>55$ would rule out a piece of the parameter space that is otherwise consistent with the WMAPII results.  Finally, we remind the reader that while it may be significant when a model fails to produce the requisite amount of inflation, if $N\gg55$ we cannot rule the model out of consideration,  as inflation can halt when an instability in the scalar field potential allows the field point to roll in an orthogonal direction. 

The running of the spectral index, $dn_s /d\ln{k}$, is displayed in  Figure \ref{fig:pl7}. This recapitulates the well known result that the running is typically very small for simple inflationary models.   If  this parameter really is significant (as is tentatively suggested by the WMAPII dataset), it will rule out essentially all inflationary models which can be described in terms of just two slow roll parameters. At the very least we would need to include higher order terms in the slow roll expansion in order to fit such a signal.   Further, the large reduction is the allowed values of the running when one adds the SDSS to the fit (as seen in Figure \ref{fig:pl4}) is explained here, since the 95\% contour for WMAPII corresponds to a much larger value (although still tiny) of $dn_s/d\ln{k}$ than the corresponding contour for WMAPII+SDSS - which follows from the quadratic dependence of $dn_s/d\ln{k}$ on $\epsilon$ and $\eta$.   Finally,  Figure \ref{fig:pl8} shows the division of the inflationary parameter space into large field, small field and hybrid models \cite{Dodelson:1997hr,Kinney:1998md}, overlaid with our constraints on $\epsilon$ and $\eta$.  We see that all three classes of models are allowed at the $1\sigma$ level, but the range of parameter values within each class is tightly restricted by the WMAPII+SDSS results.

It is of particular interest to us that the differences between the constraints found from WMAPII and WMAPII+SDSS appear to arise via the dependence of $dn_s/d\ln{k}$ on $\epsilon$ and $\eta$.  In particular, one can see from Figure~\ref{fig:pl7} that the degeneracy direction in the $(\epsilon,\eta)$ plane defined by constant $n_s$ (at lowest order in slow roll) is broken when one adds the running terms.  Fortuitously,  as one moves out along this line, $dn_s/d\ln{k} > 0$ and $r$ grows with $\epsilon$. In a conventional fit, we see a degeneracy between the tensor amplitude, the scalar amplitude and the scalar spectral index. Increasing $r$ adds power to the $C_\ell$ in the Sachs-Wolfe tail of the spectrum, and $A_s$ is driven down to compensate. However, the tensors do not contribute to the temperature anisotropies past the first acoustic peak and to large structure, so $n_s$ rises to offset the lower scalar amplitude at small scales. 

By using the HSR equations, we necessarily increase $dn_s/d\ln{k}$ when we increase $\epsilon$ and $r$ --  and the ``lever arm'' felt by this term is magnified when we include SDSS data.   The HSR approach thus implicitly imposes a ``higher order'' consistency condition, where the running of the scalar index is correlated with the tensor amplitude    \cite{Copeland:1993zn,Liddle:1994cr}.  To our knowledge, this is the first time that a cosmological dataset has been detailed enough to justify the inclusion of such higher order correlations, underscoring the importance of the new WMAPII results.   We would add two qualifications to this claim however.  The first is that we only work with two slow roll parameters. In the usual expansion $dn_s/d\ln{k}$ depends on $H'''(\phi)$, so the results may change significantly if we add the $\xi$ term to our parameter set.\footnote{Or, equivalently, $V'''(\phi)$.}  Secondly, while we know from Spergel {\em et al.\/}  \cite{Spergel:2006hy} that the data is consistent with a spectral index that decreases with  $k$, a mildly positive running is ruled out.  If the central value for the running was closer to zero, the ability of these higher order terms to constrain the inflationary parameter space would be reduced. On the other hand, if $dn_s/d\ln{k}$ is close to the central value suggested by \cite{Spergel:2006hy}, our results would imply that theories based on a single, minimally coupled inflaton is inconsistent with observations if the potential can be well specified solely in terms of $\epsilon$ and $\eta$, ruling out a vast class of simple inflationary models. 

\section{Conclusions \label{sec:conclude}}

In this paper we have introduced a new approach to introducing an inflationary prior into a cosmological parameter estimation. We proceed by varying the values of the the Hubble Slow Roll parameters directly inside the Monte Carlo Markov Chains, and evolving the values of these parameters using the ``flow equations'' of the slow roll hierarchy.  This approach has much in common with Leach {\em et al.\/}'s use of the slow roll expansion in cosmological parameter estimations  \cite{Leach:2002ar,Leach:2002dw,Leach:2003us}, and Monte Carlo reconstruction which uses the flow hierarchy to generate inflationary models which match a predefined set of constraints \cite{easther/kinney:2003}, combining the advantages of both techniques. In particular, since we do not need to expand the power spectrum or other physical quantities as a Taylor series about some fixed point, we expect that this method will be particularly suited to high quality, heterogeneous datasets that we expect will become available  in the future.   

We can see several directions in which to extend this work. The weak but persistent evidence for a large running in the scalar spectrum found in the WMAPII data provides a compelling motivation for extending our results to include $\xi$, the third order term in the slow roll expansion.  From a numerical perspective, this makes the likelihood surface significantly more complicated and increases the time taken for the  MCMC chains to converge, but we expect that the problem will be computationally tractable.  Moreover, as further cosmological datasets emerge, it will obviously be of considerable interest to see how the allowed region of HSR parameter space evolves and, hopefully, contracts. From a theoretical standpoint, we plan to look more carefully at the use of the HSR flow code with  high quality heterogeneous datasets that incorporate Lyman-$\alpha$ forest or other short wavelength probes of the primordial spectrum.  In particular, this approach would be very well suited to a dataset that contained a direct detection of primordial gravitational waves from BBO or a similar experiment, and we suspect this would allow us to put exquisite constraints on the form of the inflationary potential. 

On a practical level, we have used the Hubble Slow Roll equations to apply an inflationary prior to both the WMAPI+All dataset, a compendium of several CMB experiments, and the new WMAPII data, with and without the SDSS power spectrum.   In doing so, we find that the central value of $n_s$ moves slightly towards the scale free Harrison-Zel'dovich result. The discrepancy between this and the results in \cite{Spergel:2006hy} is explained by the use of slightly different assumptions in our likelihood function, and sheds light on the possible systematic errors that arise when fitting cosmological parameters to this accuracy.  There are other small systematic effects which can move the centroid of the $n_s$ distribution towards redder values, and in particular, a stronger upper limit on $\tau$ (e.g. by using a larger complement KaQVW of WMAP bands in the polarization analysis) will strengthen the case for deviation from scale invariance by cutting down the residual $n_s, \tau$ degeneracy, as will dropping the assumption of instantaneous reionization (David Spergel, private communication). This is clearly a question which more data will settle. On the other hand, we find a tighter constraint on the tensor amplitude, $r$, than provided by \cite{Spergel:2006hy}.

All of the major classes of single field inflationary models are allowed by our constraints, although many specific models are ruled out. However, we find evidence that the combination WMAPII+SDSS is powerful enough to be sensitive to terms that are second order in the slow roll parameters $\epsilon$ and $\eta$, via the contribution of these terms to the running of the scalar spectral index, $dn_s/ d\ln{k}$.  This evidence is tentative and  could easily evaporate as the overall cosmological dataset continues to improve, but it is indicative of the usefulness of the Hubble slow roll formalism as a tool for analyzing cosmological data. Finally, and independently of our detailed results, we can conclude that the era of precision cosmology is now in full flight, and that the promise that astrophysical data will put meaningful constraints on the inflationary potential is beginning to be fulfilled. 

\acknowledgments{We thank Antony Lewis for guidance in modifying CosmoMC, and are grateful to  Dragan Huterer, Will Kinney, Eugene Lim, and members of the WMAP science team for useful discussions. We wish to thank the staff at Yale's High Performance Computing facility for their assistance. RE is supported in part by the United States Department of Energy, grant DE-FG02-92ER-40704.  HVP is supported by NASA through Hubble Fellowship grant \#HF-01177.01-A awarded by the Space Telescope Science Institute, which is operated by the Association of Universities for Research in Astronomy, Inc., for NASA, under contract NAS 5-26555.}


\begin{thebibliography}{99}

\bibitem{lidsey/etal:1995}
  J.~E.~Lidsey, A.~R.~Liddle, E.~W.~Kolb, E.~J.~Copeland, T.~Barreiro and M.~Abney,
  Rev.\ Mod.\ Phys.\  {\bf 69}, 373 (1997)
  [arXiv:astro-ph/9508078].


\bibitem{Lyth:2001nq}
  D.~H.~Lyth and D.~Wands,
  Phys.\ Lett.\ B {\bf 524}, 5 (2002)
  [arXiv:hep-ph/0110002].
  
  

\bibitem{hoffman/turner:2001}
  M.~B.~Hoffman and M.~S.~Turner,
  Phys.\ Rev.\ D {\bf 64}, 023506 (2001)
  [arXiv:astro-ph/0006321].

\bibitem{kinney:2002}
  W.~H.~Kinney,
  Phys.\ Rev.\ D {\bf 66}, 083508 (2002)
  [arXiv:astro-ph/0206032].
 


\bibitem{Copeland:1993zn}
  E.~J.~Copeland, E.~W.~Kolb, A.~R.~Liddle and J.~E.~Lidsey,
  Phys.\ Rev.\ D {\bf 49}, 1840 (1994)
  [arXiv:astro-ph/9308044].
  
\bibitem{Liddle:1994cr}
  A.~R.~Liddle and M.~S.~Turner,
  Phys.\ Rev.\ D {\bf 50}, 758 (1994)
  [Erratum-ibid.\ D {\bf 54}, 2980 (1996)]
  [arXiv:astro-ph/9402021].



\bibitem{easther/kinney:2003}
  R.~Easther and W.~H.~Kinney,
  Phys.\ Rev.\ D {\bf 67}, 043511 (2003)
  [arXiv:astro-ph/0210345].
  

  
\bibitem{Leach:2002ar}
  S.~M.~Leach, A.~R.~Liddle, J.~Martin and D.~J.~Schwarz,
  Phys.\ Rev.\ D {\bf 66}, 023515 (2002)
  [arXiv:astro-ph/0202094].
  
\bibitem{Leach:2002dw}
  S.~M.~Leach and A.~R.~Liddle,
  Mon.\ Not.\ Roy.\ Astron.\ Soc.\  {\bf 341}, 1151 (2003)
  [arXiv:astro-ph/0207213].
  
\bibitem{Leach:2003us}
  S.~M.~Leach and A.~R.~Liddle,
  Phys.\ Rev.\ D {\bf 68}, 123508 (2003)
  [arXiv:astro-ph/0306305].


\bibitem{Martin:2000ak}
  J.~Martin, A.~Riazuelo and D.~J.~Schwarz,
  Astrophys.\ J.\  {\bf 543}, L99 (2000)
  [arXiv:astro-ph/0006392].

\bibitem{Stewart:2001cd}
  E.~D.~Stewart,
  Phys.\ Rev.\ D {\bf 65}, 103508 (2002)
  [arXiv:astro-ph/0110322].

\bibitem{Ungarelli:2005qb}
  C.~Ungarelli, P.~Corasaniti, R.~A.~Mercer and A.~Vecchio,
  Class.\ Quant.\ Grav.\  {\bf 22}, S955 (2005)
  [arXiv:astro-ph/0504294].

\bibitem{Boyle:2005se}
  L.~A.~Boyle and P.~J.~Steinhardt,
  arXiv:astro-ph/0512014.


\bibitem{Smith:2006xf}
T.~L.~Smith, H.~V.~Peiris and A.~Cooray,
arXiv:astro-ph/0602137.


\bibitem{Chongchitnan:2006pe}
S.~Chongchitnan and G.~Efstathiou,
arXiv:astro-ph/0602594.



\bibitem{Lidz:2003fv}
  A.~Lidz, L.~Hui, A.~P.~S.~Crotts and M.~Zaldarriaga,
  arXiv:astro-ph/0309204.

\bibitem{McQuinn:2005hk}
  M.~McQuinn, O.~Zahn, M.~Zaldarriaga, L.~Hernquist and S.~R.~Furlanetto,
  arXiv:astro-ph/0512263.

\bibitem{Abbott:1984fp}
  L.~F.~Abbott and M.~B.~Wise,
  Nucl.\ Phys.\ B {\bf 244}, 541 (1984).

\bibitem{Easther:1995pc}
  R.~Easther,
  Class.\ Quant.\ Grav.\  {\bf 13}, 1775 (1996)
  [arXiv:astro-ph/9511143].
  
\bibitem{Starobinsky:2005ab}
  A.~A.~Starobinsky,
  JETP Lett.\  {\bf 82}, 169 (2005)
  [Pisma Zh.\ Eksp.\ Teor.\ Fiz.\  {\bf 82}, 187 (2005)]
  [arXiv:astro-ph/0507193].


\bibitem{Grivell:1999wc}
  I.~J.~Grivell and A.~R.~Liddle,
  Phys.\ Rev.\ D {\bf 61}, 081301 (2000)
  [arXiv:astro-ph/9906327].



\bibitem{Kinney:2005in}
  W.~H.~Kinney and A.~Riotto,
  arXiv:astro-ph/0511127.

\bibitem{Lyth:1996im}
  D.~H.~Lyth,
  Phys.\ Rev.\ Lett.\  {\bf 78}, 1861 (1997)
  [arXiv:hep-ph/9606387].
  
\bibitem{Easther:2006qu}
  R.~Easther, W.~H.~Kinney and B.~A.~Powell,
  arXiv:astro-ph/0601276.

\bibitem{Dimopoulos:2005ac}
  S.~Dimopoulos, S.~Kachru, J.~McGreevy and J.~G.~Wacker,
  arXiv:hep-th/0507205.
  
\bibitem{Easther:2005zr}
  R.~Easther and L.~McAllister,
  arXiv:hep-th/0512102.


  
\bibitem{peiris/etal:2003}
  H.~V.~Peiris {\it et al.},
  Astrophys.\ J.\ Suppl.\  {\bf 148}, 213 (2003)
  [arXiv:astro-ph/0302225].


\bibitem{Adams:2001vc}
  J.~A.~Adams, B.~Cresswell and R.~Easther,
  Phys.\ Rev.\ D {\bf 64}, 123514 (2001)
  [arXiv:astro-ph/0102236].
 
\bibitem{Spergel:2003cb}
  D.~N.~Spergel {\it et al.}  [WMAP Collaboration],
  Astrophys.\ J.\ Suppl.\  {\bf 148}, 175 (2003)
  [arXiv:astro-ph/0302209].

\bibitem{Bennett:2003bz}
  C.~L.~Bennett {\it et al.},
  Astrophys.\ J.\ Suppl.\  {\bf 148}, 1 (2003)
  [arXiv:astro-ph/0302207].

\bibitem{Kuo:2002ua}
  C.~l.~Kuo {\it et al.}  [ACBAR collaboration],
  Astrophys.\ J.\  {\bf 600}, 32 (2004)
  [arXiv:astro-ph/0212289].

\bibitem{Mason:2002tm}
  B.~S.~Mason {\it et al.},
  Astrophys.\ J.\  {\bf 591}, 540 (2003)
  [arXiv:astro-ph/0205384].

\bibitem{Grainge:2002da}
  K.~Grainge {\it et al.},
  Mon.\ Not.\ Roy.\ Astron.\ Soc.\  {\bf 341}, L23 (2003)
  [arXiv:astro-ph/0212495].

\bibitem{Freedman:2000cf}
  W.~L.~Freedman {\it et al.},
  Astrophys.\ J.\  {\bf 553}, 47 (2001)
  [arXiv:astro-ph/0012376].
  
  
\bibitem{Hinshaw:2006ia}
  G.~Hinshaw {\it et al.},
  arXiv:astro-ph/0603451.
  
\bibitem{Page:2006hz}
  L.~Page {\it et al.},
  arXiv:astro-ph/0603450.
  
\bibitem{Spergel:2006hy}
  D.~N.~Spergel {\it et al.},
  arXiv:astro-ph/0603449.
 
\bibitem{kinney/etal:2004}
  W.~H.~Kinney, E.~W.~Kolb, A.~Melchiorri and A.~Riotto,
  Phys.\ Rev.\ D {\bf 69}, 103516 (2004)
  [arXiv:hep-ph/0305130].

\bibitem{liddle:2003}
  A.~R.~Liddle,
  Phys.\ Rev.\ D {\bf 68}, 103504 (2003)
  [arXiv:astro-ph/0307286].


\bibitem{Christensen:2000ji}
  N.~Christensen and R.~Meyer,
  arXiv:astro-ph/0006401.
  
\bibitem{Christensen:2001gj}
  N.~Christensen, R.~Meyer, L.~Knox and B.~Luey,
  Class.\ Quant.\ Grav.\  {\bf 18}, 2677 (2001)
  [arXiv:astro-ph/0103134].

\bibitem{Knox:2001fz}
  L.~Knox, N.~Christensen and C.~Skordis,
  arXiv:astro-ph/0109232.
  
\bibitem{Lewis:2002ah}
  A.~Lewis and S.~Bridle,
  Phys.\ Rev.\ D {\bf 66}, 103511 (2002)
  [arXiv:astro-ph/0205436].
  
\bibitem{Kosowsky:2002zt}
  A.~Kosowsky, M.~Milosavljevic and R.~Jimenez,
  Phys.\ Rev.\ D {\bf 66}, 063007 (2002)
  [arXiv:astro-ph/0206014].
  
\bibitem{Verde:2003ey}
  L.~Verde {\it et al.},
  Astrophys.\ J.\ Suppl.\  {\bf 148}, 195 (2003)
  [arXiv:astro-ph/0302218].
  
\bibitem{Caldwell:1999ew}
  R.~R.~Caldwell,
  Phys.\ Lett.\ B {\bf 545}, 23 (2002)
  [arXiv:astro-ph/9908168].
  
  
\bibitem{Khoury:2001wf}
  J.~Khoury, B.~A.~Ovrut, P.~J.~Steinhardt and N.~Turok,
  Phys.\ Rev.\ D {\bf 64}, 123522 (2001)
  [arXiv:hep-th/0103239].
  
\bibitem{Gasperini:1992em}
  M.~Gasperini and G.~Veneziano,
  Astropart.\ Phys.\  {\bf 1}, 317 (1993)
  [arXiv:hep-th/9211021].
  
\bibitem{Tegmark:2003uf}
  M.~Tegmark {\it et al.}  [SDSS Collaboration],
  Astrophys.\ J.\  {\bf 606}, 702 (2004)
  [arXiv:astro-ph/0310725].

  
\bibitem{gelman/rubin:1992}
Gelman, A. and Rubin, D. Statistical Science, {\bf 7}, 457 (1992)

\bibitem{Kadota:2005hv}
  K.~Kadota, S.~Dodelson, W.~Hu and E.~D.~Stewart,
  Phys.\ Rev.\ D {\bf 72}, 023510 (2005)
  [arXiv:astro-ph/0505158].

\bibitem{Dodelson:1997hr}
  S.~Dodelson, W.~H.~Kinney and E.~W.~Kolb,
  Phys.\ Rev.\ D {\bf 56}, 3207 (1997)
  [arXiv:astro-ph/9702166].
  
\bibitem{Kinney:1998md}
  W.~H.~Kinney,
  Phys.\ Rev.\ D {\bf 58}, 123506 (1998)
  [arXiv:astro-ph/9806259].
  
\end{thebibliography}
\end{document}